\documentclass[fleqn,usenatbib]{mnras}

\usepackage{newtxtext,newtxmath}
\usepackage[table,xcdraw]{xcolor}
\usepackage{caption}
\usepackage{soul}
\usepackage{placeins}

\usepackage[T1]{fontenc}
\setulcolor{red} 

\DeclareRobustCommand{\VAN}[3]{#2}
\let\VANthebibliography\thebibliography
\def\thebibliography{\DeclareRobustCommand{\VAN}[3]{##3}\VANthebibliography}

\usepackage{graphicx}	
\usepackage{amsmath}	
\usepackage{color}

\newcommand{\wsclean}[1]{\texttt{WSClean}}
\newcommand{\brender}[1]{\texttt{BLOB-RENDER}}

\title[Simulations of relativistic ejecta from BHs]{Relativistic ejecta from stellar mass black holes: insights from simulations and synthetic radio images}

\author[K. Savard et al.]{
Katie Savard,$^{1}$\thanks{E-mail: katherine.savard@physics.ox.ac.uk}
James H. Matthews,$^{1}$
Rob Fender$^{1,2}$
and Ian Heywood$^{1,3,4}$
\\
$^{1}$Astrophysics, Department of Physics, University of Oxford, Keble Road, Oxford, OX13RH, UK\\
$^{2}$Department of Astrophysics, University of Cape Town, Private Bag X3, Rondebosch, Cape Town 7701, South Africa\\
$^{3}$South African Radio Astronomy Observatory, 2 Fir Street, Observatory 7925, South Africa \\
$^{4}$Department of Physics and Electronics, Rhodes University, PO Box 94, Makhanda 6140, South Africa \\
}

\date{\today}
\pubyear{2015}

\begin{document}
\label{firstpage}
\pagerange{\pageref{firstpage}--\pageref{lastpage}}
\maketitle

\begin{abstract}
\noindent We present numerical simulations of discrete relativistic ejecta from an X-ray binary (XRB) with  initial
conditions directly informed by observations. XRBs have been observed to launch powerful discrete plasma ejecta during state transitions, which can propagate up to parsec distances. Understanding these ejection events unveils new understanding of jet-launching, jet power, and jet-ISM interaction among other implications. Multi-frequency quasi-simultaneous radio observations of ejecta from the black hole XRB MAXI J1820+070 produced both size and calorimetry constraints, which we use as initial conditions of a relativistic hydrodynamic simulation. We qualitatively reproduce the observed deceleration of the ejecta in a homogeneous interstellar medium (ISM). Our simulations demonstrate that the ejecta must be denser than the ISM, the ISM be significantly low-density, and the launch be extremely powerful, in order to propagate to the observed distances. The blob propagates and clears out a high-pressure low-density cavity in its wake, providing an explanation for this pre-existing low-density environment, as well as ‘bubble-like’ environments in the vicinity of XRBs inferred from other studies. As the blob decelerates, we observe the onset of instabilities and a long-lived reverse shock – these mechanisms convert kinetic to internal energy in the blob, responsible for in-situ particle acceleration.
We transform the outputs of our simulation into pseudo-radio images, incorporating the $u$,$v$ coverage of the MeerKAT and e-MERLIN telescopes from the original observations with real-sky background. Through this, we maximize the interpretability of the results and provide direct comparison to current data, as well as provide prediction capabilities.

\end{abstract}




\section{Introduction}\label{sec:intro}

The long timescales and vast distances associated with most astronomical phenomena arguably pose the greatest challenges to observational astronomy. X-ray binaries (XRBs), however, manage to sit in an observable sweet-spot, as they are amongst the relatively few astrophysical systems in the Universe that are resolvable in both space and time. They provide fantastic physics laboratories that allow us to study the extremes of gravitation, accretion, and high energy physics with precision.

XRBs consist of compact objects (black hole or neutron star) that are in binary orbit with a companion star, close enough that through Roche-lobe overflow or strong winds (or both), material is stripped from the star forming an accretion disk about the compact object \citep[see][for reviews]{tanaka1996,done2007everything,bahramian_review}. They are known to launch powerful extended jets transiently throughout their lifetime, but the exact origin, composition, and nature of these jets is still debated.

Black hole XRBs with low-mass companion stars (LMXRBs) usually spend the majority of their life in a very low luminosity `quiescent' state, but are observed, and frequently discovered, when they go into outburst. Overall, the lifetime of an outburst consists of traversing hysteretically between states of accretion,  intrinsically linked to jet launching \citep{fender2004towards,belloni2012states,done2007everything, dunn2010global}. On one end, XRBs can occupy the `Hard' state, where compact self-absorbed synchrotron emitting radio jets are produced and a hard non-thermal spectrum, associated with hot coronal emission, is observed in the X-rays \citep{fender2001powerful}. On the other end, the `Soft' state is a high-luminosity state where the compact jet is quenched and the thermal emission from the disk dominates the X-rays, resulting in a softer spectrum \citep{fender1999quenching}. Some of this behaviour also extends to XRBs with high-mass companions and/or neutron star accretors \citep{munoz-darias2014}, although the phenomenology is clearer and more typical for BH LMXRBs.

A general description of the coupling between the jet and the accretion disk, specifically the way XRBs evolve between canonical states, was originally presented by \cite{fender2004towards}. In this description, XRBs follow a cycle in X-ray hardness and X-ray intensity (now ubiquitously the hardness-intensity diagram, HID). An interpretation of this behaviour is that the X-ray hardness correlates with the inner-radius of the disk and the existence of a jet (where jet Lorentz factor is likely maximum at the transition), thus linking the accretion state with jet launching. By far the most observationally striking transition occurs between the hard and soft state, where the source brightens in the hard state and as the spectrum softens we observe a bright flare of (typically) optically-thick synchrotron emission and sometimes (perhaps always, but unconfirmed observationally) the launch of transient ejecta as the jet turns off. These ejecta resemble discrete blobs of optically thin synchrotron emitting plasma, as opposed to the continuous compact jet seen beforehand.

Transient XRB ejecta had been identified in some sources by the early 1990s, most notably in SS 433 \citep{hjellming1981analysis} and Cyg X-3 \citep{molnar1988vlbi}, where "radio bubbles" were observed with inferred outflow velocities of $\sim0.2$c. However, it was not until the observations of the BH LMXRB GRS 1915+105 in outburst by \cite{mirabel1994superluminal} that apparent superluminal motion within our Galaxy was first confirmed. \citet{mirabel1994superluminal} suggest that some central compact object is `ejecting matter in a process similar to, but on a smaller scale than that seen in quasars', making the essential link between understanding relativistic ejections from XRBs as a proxy for understanding jets on larger and more distant scales. Following this discovery, dozens of other examples of these resolved discrete ejecta were observed \citep{hjellming1995episodic,mioduszewski2001,gallo2004transient}, including repeated ejections from GRS 1915+105 \citep{fender1999merlin, dhawan2000synchrotron}. 

Observations of radio and X-ray jets from XTE J1550-564 \citep{corbel2002large} suggested that at least some BH XRB jets were powerful enough to propagate to parsec scales before decelerating, but until recently it was not clear how common this phenomenon was. What was clear was that if deceleration on large-scales could be regularly tracked, it could lead to a new, more precise, measurements of jet powers and properties. Recently, the ThunderKAT collaboration \citep{2016mks..confE..13F} has revolutionized the study of transient jets through monitoring and follow-up campaigns with the MeerKAT telescope, observing these ejecta decelerating over unprecedented spatial scales (e.g. \citealt{russell2019disk, Bright2020, carotenuto2022modelling, bahramian2023maxi}).

These events provide a unique and rich laboratory for studying relativistic outflows. For instance, the kinematics and luminosity of these ejecta can potentially help constrain our understanding of: the material of the jet, the kinetic power of the jet, particle acceleration mechanisms in the jet, as well as the transfer of energy and dissipation mechanisms between the jet and the ISM. 

Recently, powerful superluminal ejecta were launched from the BH XRB MAXI J1820+070, and radio \citep{Bright2020} and X-ray \citep{espinasse2020} observations revealed their trajectories over parsec scales. In particular, a unique estimate of the internal energy of the plasma `blob' ejected from the XRB was made 90 days after launch, along with a constraint on the size of the blob. This, in conjunction with the mapping of the deceleration and flux as a function of time, provides groundwork for understanding the kinematics and smaller-scale shock physics of relativistic ejecta, and opens up doors to exploring these systems from a numerical perspective.

A great deal of work has been done through simulations and modelling to understand the physics of jets, much of it focusing on the question of how jets are launched. The most common descriptions of jet launching rely on the presence of large-scale poloidal magnetic fields which serve to extract rotational energy and angular momentum and transfer it to the plasma that forms the jets. The two most widely referenced types of jet models of this description are the \cite{blandford-znajek1997} and \cite{blandford-payne1982} models: the former predicts that energy and angular momentum are extracted directly from the spin of the black hole; the latter predicts that the extraction is from the disk itself. It is likely that these mechanisms can be distinguished (or even ruled out in favour of other models) with a greater understanding of jet properties such as morphology and overall power. Beyond jet launch, the stability, propagation, acceleration, collimation, and composition of jets are all active areas of study \citep[see ][for reviews]{blandford_relativistic_2019,davis_magnetohydrodynamic_2020}.

General Relativistic Magneto-Hydrodynamic (GRMHD) simulations have been at the forefront for progressing our understanding of jet physics close to the central black hole \citep[e.g.][]{mckinney2006general-relativistic,Tchekhovskoy2011,nakamura2018m87,liska2020,liska2024,Lalakos2024}, while (M)HD simulations  without GR have typically been used to study jet propagation on much larger scales, particularly in the context of active galactic nuclei \citep[e.g.][]{mignone2010highres,morsony2007temporal,hardcastle2013numerical,Hardcastle2014,Tchekhovskoy2016,matthews_ultrahigh_2019,henwy2023} and gamma-ray bursts  \citep[GRBs; e.g.][]{granot2001,meliani2007,zhang2009,vaneerten2010,vaneerten2012,ayache2022}. Simulations of jets specifically in XRBs \citep{perucho2008interaction,goodall2011microquasar,monceau2014,lopez-miralles2022simulations} have to date mostly been limited to `steady' jets \citep[although see][]{smponias2023}. With an increasing sample of high-quality observations of transient ejecta in particular, there is now an opportunity to study jets from a new perspective. 


\cite{carotenuto2022modelling} applied a propagating shock model originally developed for GRBs by \citet{wang2003external} to model the kinematics of ejected blobs from the BH LMXRB MAXI J1348-630. In particular, the final deceleration phase allowed \citet{carotenuto2022modelling} to place constraints on the jet Lorentz factor, inclination angle, as well as internal and kinetic energy. In addition, information about the density structure of the surrounding ISM was obtained through this modelling. \cite{carotenuto2022modelling} infer that the blob travelled through an under-dense ISM cavity before decelerating as it entered the denser cavity edge. This low density environment is consistent with other observations and models \citep{heinz2002radio-lobe,hao2009large-scale} that propose most BH XRBs are found in low-density cavity environments, perhaps due to previous jet activity in the hard state or accretion disk winds. A similar study of the LMXRB XTE J155-560 by \cite{migliori2017evolving} consisting of both radio and X-ray observations provides strong evidence for the existence of an ISM cavity as well, where particle acceleration and radiative dissipation is theorized to be triggered by the ejecta impacting the jump in density gradient. These cavity edges are therefore probable sites of major jet/ISM interaction, and significant amounts of energy can be dissipated through various interaction mechanisms \citep{goodall2011microquasar, tetarenko2018mapping}. Exactly how energy is transferred between the jet and the ISM is not well understood, although the common framework is through shocks and turbulent mixing \citep{bromberg2011,kusafuka2023}. 

It is clear that simulations of these discrete ejecta have the potential to explain how these blobs decelerate, disrupt, and therefore transfer energy into the ISM. They also have the capacity to constrain parameters like density (both absolute and contrasting), kinetic and internal energy, and potentially particle acceleration efficiency. Through this, we not only progress our understanding of jet physics but also of ISM density profiles and feedback mechanics, with implications for galactic energy injection and possible star formation (\citealt*{fender2005energization};\citealt{mirabel2014jet}).

In this paper, we present simulations of discrete jet ejecta, or blobs, designed to emulate those observed in XRBs such as MAXI J1820+070. We begin by describing our relativistic hydrodynamic simulations and other aspects of methodology in section~\ref{methods}. We present our results in section~\ref{simresults}, and describe our new \texttt{BLOB-RENDER} software in section~\ref{ezblob}. We discuss our results and the wider implication in section~\ref{discussion}, before concluding in section~\ref{conclusions}. 

\section{Methods}
\label{methods}

In this study, we perform simulations of relativistic discrete ejecta from XRBs with initial conditions directly informed by observations. Specifically, we use the physical conditions inferred from the radio observation of the relativistic ejecta from MAXI J1820+070 \citep{Bright2020} 90 days after the detection of a radio flare likely associated with the launch of the ejecta. The details of this measurement, and how they are implemented as numerical initial conditions, are outlined in section \ref{initialconditions}. Physical parameters of the ejecta/ISM interaction such as density contrast, absolute density, particle acceleration efficiencies, etc. are largely unconstrained from the observations alone. However, these parameters can be tuned in simulations in order to discern their effect on both the kinematics of the ejecta and its lightcurve, and in turn compared to the real lightcurve and kinematic data obtained after the 90-day measurement.

\subsection{Numerical methods}\label{nummethods}

The simulations in this paper are numerical solutions to the special relativistic hydrodynamic (RHD) system of equations, performed using the \textsc{Pluto} code \citep{Mignone2007}. The numerical scheme evolves the conservative set of variables 
\begin{center}
\begin{math}
    \boldsymbol{U} = (D,m_1,m_2,m_3,{\cal E}_t)^T
\end{math}
\end{center}

\noindent with laboratory mass density $D$, momenta $m_{1,2,3}$ and total energy ${\cal E}_t$ densities,  according to the conservation equations

\begin{center}
\begin{math}
\frac{\partial}{\partial t}\begin{pmatrix}
    D \\ m \\ {\cal E}_t
\end{pmatrix} + \nabla \cdot \begin{pmatrix}
    D\boldsymbol{v} \\ \boldsymbol{mv} + P {\cal I} \\ \boldsymbol{m}
\end{pmatrix}^T = \begin{pmatrix}
    0 \\ 0 \\ 0
\end{pmatrix}
\end{math}
\end{center}

\noindent with velocity $\boldsymbol{v}$, thermal pressure $P$ and identity matrix ${\cal I}$. 
The relation between the conserved variables $\boldsymbol{U}$ and the primitive variables is expressed as
\begin{center}
$D = \rho \Gamma$ \\
$\boldsymbol{m} = \rho h \Gamma^2 \boldsymbol{v}$   \\
${\cal E}_t = \rho h \Gamma^2 - P$ 
\end{center}
where $\rho$ is the rest-mass density, $\Gamma$ the Lorentz factor, and
$h$ the specific enthalpy. We use the Taub-Matthews equation of state \citep{taub_relativistic_1948,mignone_equation_2007}, describing a relativistic perfect gas, which is implemented through a quadratic approximation. The adiabatic exponent in this equation of state is a function of temperature, and therefore ranges from $\gamma_{\rm ad}=5/3$ to $\gamma_{\rm ad}=4/3$ in the low and high temperature limits respectively, allowing adaptation to both relativistic and non-relativistic regimes. 

We require the RHD \texttt{PLUTO} module due to the significant Lorentz factors of the ejecta. As the distance to the compact object in these simulations is sufficiently large ($>10^{10} r_g$), general relativistic effects are negligible. In this description, we exclude magnetic fields for several reasons. Firstly, the effect of small-scale magnetic fields on emission mechanisms such as synchrotron radiation can be inferred by parametrizing the partitioning between particle and magnetic field energy density. Specifically, we assume the minimum energy scenario, both due to plausibility and simplicity. Secondly, both the strength and structure of large-scale magnetic fields which could have dynamical impacts on the ejecta are currently poorly understood, and for this treatment we opt to exclude them to avoid erroneous assumptions, although we discuss their possible impact in section \ref{discussion}.

The simulations in this work were run in 2D cylindrical coordinates, with the exception of the 3D and 1D test simulations, discussed separately in section~\ref{3dsims} and in \ref{instabilities_shocks}. This geometry lends itself well to the expected natural geometry of the ejecta. We use a linear reconstruction of primitive variables as the spatial order of integration, and a $2^{\rm nd}$ order total variation diminishing (TVD) Runge-Kutta scheme to evolve timesteps, with a Courant-Friedrichs-Lewy (CFL) number of $C_a=0.2$ in 2D and $C_a=0.15$ in 3D. For flux computation, we use the Harten–Lax–van Leer contact (HLLC) Riemann solver \citep{mignone_hllc_2005}, the least diffusive solver available in \textsc{Pluto} for RHD equations that isn't inhibited by significant computational cost.  In addition, we employ the least diffusive flux limiter available (monotized central difference).

\begin{figure}
    \centering
    \includegraphics[width=0.9\columnwidth]{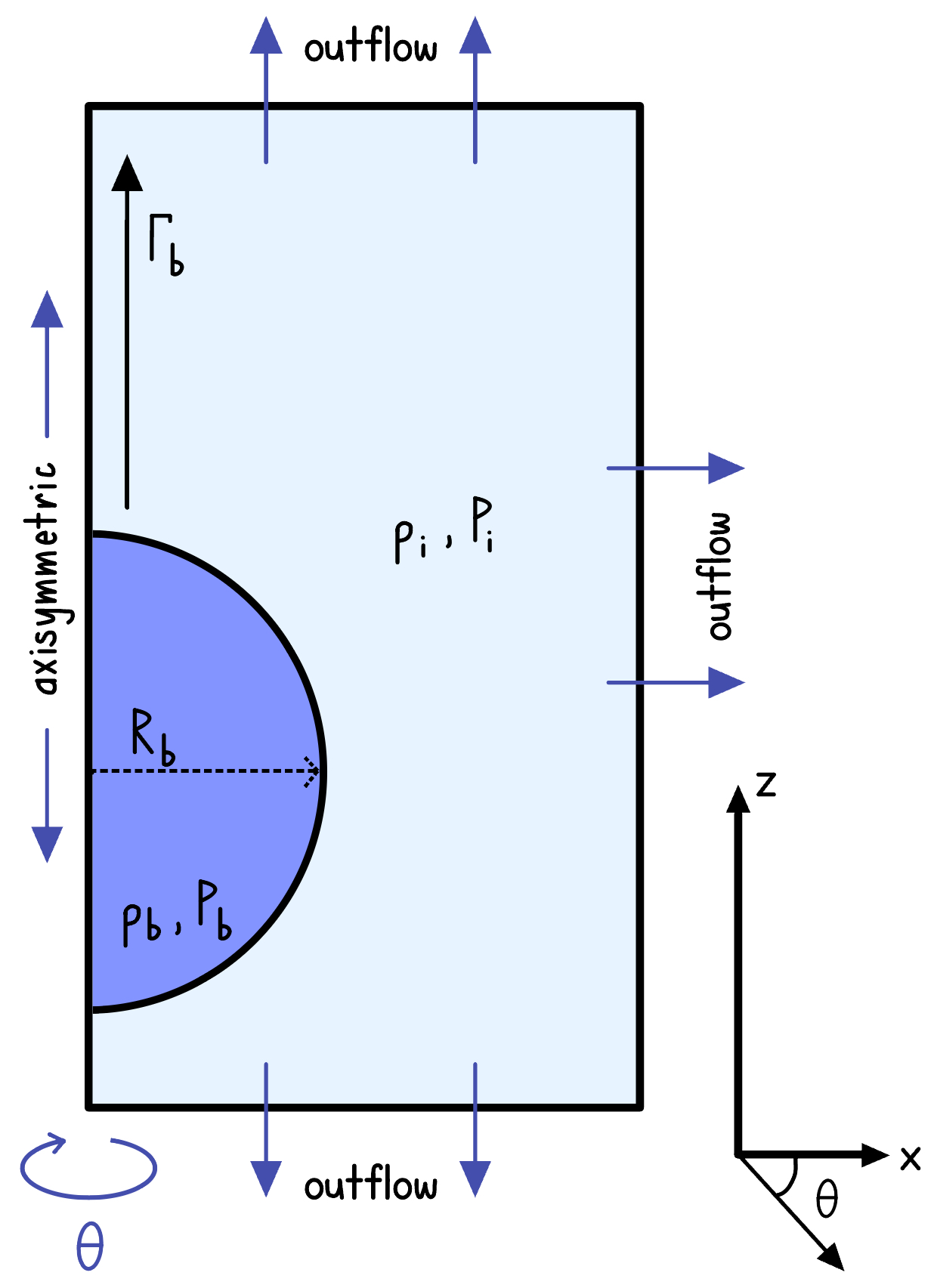}
    \caption{Diagram of the \textsc{Pluto} initial conditions. Simulations are run in 2D cylindrical coordinates: the domain is rotated about the z axis and each cell is square toroidal in shape. Density and pressure are specified inside ($\rho_b,P_b$) and outside ($\rho_i,P_i$) of the radius $R_{b}$ which defines the border between the blob and ISM. Initial condition values are listed in Table \ref{Tab:table_fid}.}
    \label{fig:diagram}
\end{figure}

We split the 2D domain into $3000\times6000$ computational cells, with each simulation unit 6 cells across, equivalent to a 2D domain span of $500\times1000$ simulation units. This 2D uniform grid is rotated fully about its y-axis such that each cell is square toroidal in shape and the computational domain is 2D cylindrical. All but the axial (axisymmetric) boundaries are set with outflowing boundary conditions. See Figure \ref{fig:diagram} for a diagram of the simulation geometry. The 3D simulation grid is made up of $800\times1500\times800$ cartesian computational cells, domain size of $400\times 750\times 400$ simulation units.
The units in \textsc{Pluto} are non-dimensional (`code units') to avoid working with overly large or small numbers which may lead to excessive numerical error. Three fundamental scaling units must be defined: unit density $\rho_0$, unit length $L_0$, and unit velocity $v_0$, for which we adopt: $\rho_0=1.673\times 10^{-24}\,\text{g cm}^{-3}$, $L_0=10^{15}$ cm and $v_0=c$, making units of time $t_0=1.058\times10^{-3}$ years, or $3.34\times10^{4}$ seconds. Each simulation cell has a physical resolution of $0.1667\times 10^{15}$cm. These values are chosen to be appropriate for the scales of an XRB jet ejection (as informed in particular by recent MeerKAT observations). We adjusted the resolution of the simulations such that those at higher resolutions than the fiducial produced near-identical deceleration and luminosity profiles. The resolution was marginally increased beyond this in order to better identify and track the development of shocks in the simulation.

\subsection{Initial conditions}\label{initialconditions}

The initial conditions of the simulation are informed directly from the observations of ejecta from MAXIJ1820+070 by \citet{Bright2020} on the 5th-7th October 2018, 90 days after a $\sim 46$mJy flare event detected from the BHXRB, observed with the Arcminute Microkelvin Image Large Array (AMI-LA). Quasi-simultaneously, MeerKAT (1.28 GHz) and eMERLIN (1.51 GHz) observations were obtained. Both telescopes operate within very similar frequencies but probe different angular scales, meaning that comparing both images can further constrain the spatial distribution of the source flux on different angular scales. With a synthesized beam of $7.9\arcsec\times 5.4\arcsec$, MeerKAT detected an unresolved source of $2.1 \pm0.1$mJy, while eMERLIN detected $0.31\pm0.02$mJy of unresolved flux with a synthesized beam of $0.3\arcsec\times0.2\arcsec$. Essentially, this can be explained as 85\% of the flux detected by MeerKAT lying on scales larger than those probed by eMERLIN, but below scales accessible to be resolved by MeerKAT, thus constraining the angular size of the ejecta (although some subtleties to this are discussed in Section \ref{t0_blobrender}). The distance to the core is constrained through radio parallax as $2.96\pm0.33$kpc \citep{atri2020parallax}, constraining the physical size to $\sim 6.2\times10^{2}$ AU to $\sim1.7\times10^{4}$ AU \citep{Bright2020}. This constraint is directly incorporated in the initial size of the simulated blob, and we simulate the evolution and trajectory of the blob from 90 days onwards. 

With knowledge of both the size and luminosity of a source, we can estimate a minimum internal energy as per \citet{longair1995}, assuming a spherical emitting volume. \citet{Bright2020} calculate this as $E_{\rm int}=$ $2.1\times10^{41} $ to $ 1.5\times 10^{43}$ erg, where the range is reflective of the uncertainty on the size estimate. A lower limit to the bulk Lorentz factor $\Gamma>1.7$ was also derived from the observed kinematics, and we incorporate a conservative estimate of $\Gamma=2.0$ in the simulations (this choice is further discussed in section \ref{fidsim}). The relativistic motion results in the beaming of the emitted radiation -- this effect therefore contributes to the calculation of the minimum energy. Incorporating the relativistic effects, including the angle to the line of sight $\theta=62^{\circ}$, shifts the minimum energy up from its original value in \citet{Bright2020} to $3.3\times10^{41}$ to $2.3\times10^{43}$ erg.

In the simulation, we assign $R_{\rm blob}=9.51\times10^{15}$cm as the initial radius of the blob, corresponding to the lower end of the possible size range. The initial internal energy is then calculated with this size, the observed flux, and $\Gamma=2.0$. We discuss the conversion from internal energy to initial pressure in the simulated blob in the latter part of Section \ref{analysisandunits}. We assume a spherical blob, uniform in density, pressure, and $\Gamma$: the exact morphology of the blob can not be inferred from the observations due to insufficient $u$,$v$ coverage. All simulation initial conditions are listed in Table \ref{Tab:table_fid}, where these initial conditions correspond to $t_{\rm obs}=90$ days.

The densities of XRB ejecta and the ISM surrounding these systems are generally not well known. One aim of these simulations is to constrain these parameters. We calculate an initial upper limit estimate of the blob density, assuming it has not swept up mass since launch. The flare associated with the launch of the ejecta, and supposedly the production of the ejecta, lasted 6.7 hours. The maximum amount of mass accreted during this timeframe at the Eddington limit for a $\sim10 M_{\odot}$ BH \citep{atri2020parallax} is roughly $3\times10^{23}$ grams. If all of this mass is (impossibly) launched into the ejecta during the flare and is confined within the smallest estimated comoving volume calculated at 90 days post-launch, we infer a maximum proton number density of $\sim 0.02$ cm$^{-3}$. This was used as a guiding point for the first simulations.

\subsection{Synchrotron emissivity}\label{analysisandunits}

In order to make direct comparisons to observations, we need a treatment of synchrotron emission. We assume that the blobs emit optically thin synchrotron radiation, which we replicate with a `pseudo-emissivity' in the simulations, following, e.g., \cite{hardcastle2013numerical}. We begin by making the standard assumption that the magnetic field and electron energy densities are related by a partitioning factor, $\eta$, such that 
\begin{gather}
    U_{\rm B} =  \eta U_{\rm e}. \label{eq:equip}
\end{gather}
We then make use of the relation between the co-moving frame synchrotron emissivity $j^\prime_{\nu^\prime}$ and the magnetic field strength \citep{longair2011} 
\begin{gather}
    j^\prime_{\nu^\prime} \propto C B^{(p+1)/2} \label{eq:emissivity}
\end{gather}
where $C$ is the constant scaling term for a power-law distribution of electron energies
\begin{gather}
        N(E)dE = C E^{-p}dE
\end{gather}
and $p$ is the power-law index. By integrating, we get the electron total internal energy density
\begin{gather}
    U_{\rm e}= \int^{E_2}_{E_1} EN(E)dE = C \frac{E_2^{2-p}-E_1^{2-p}}{2-p}  ,
\end{gather}
the latter form valid for $p\neq 2$. \footnote{For $p=2$ we instead obtain $U_{\rm e}=C \ln{(E_2/E_1)}$. We set p=2.01 for ease of computation.}
From the above, $U_{\rm e} \propto C $ (for fixed $E_1$, $E_2$ and $p$), and therefore from Equation~\ref{eq:equip} $U_{\rm B}\propto C$. Substituting this into Equation \ref{eq:emissivity}, and with $B\propto U_{\rm B}^{1/2}$, we find that
\begin{gather}
    j^\prime_{\nu^\prime}\propto U_{\rm B}^{(p+5)/4}.
\end{gather}
Reformulating this in terms of pressure ($P \propto P_{\rm B} \propto U_{\rm B}$)
we can then conclude that
\begin{gather}
    j^\prime_{\nu^\prime}\propto P^{(p+5)/4}\label{eq:em_to_pressure}.
\end{gather}
Therefore for a power-law index of $p=2$ (as measured by \citealt{Bright2020}), the pseudo-emissivity in the simulation scales with pressure as $P^{1.75}$.

To convert to the physical units of, for example, mJy, we follow the prescription from \cite{hardcastle2013numerical}, such that
\begin{gather}
    j^\prime_{\nu^\prime} = A(\alpha,\nu',\gamma,\eta) (\kappa P)^{(p+5)/4}\label{eq:emis_to_prs_full}.
\end{gather}
and the constant $A$ encapsulates the dependence on $\eta$, observing frequency and electron distribution. We assume a power-law electron energy index $p=2.01$ (and as a result $\alpha\simeq-0.5$), $\eta=0.75$ (utilizing the minimum energy assumption), $\nu=1.28$~GHz (MeerKAT L band), and energy limits on the electron distribution $\gamma=10-10^7$. We then convert this to Janskys per pixel, using the radio-parallax distance of $2.96\pm0.33$ kpc measured by \cite{atri2020parallax}.

Here, we have also introduced an additional efficiency term $\kappa$, which provides a simplistic treatment for particle acceleration, delegating how much of the internal energy is stored in relativistic particles and magnetic fields. The efficiency term $\kappa$ does not correspond directly to the particle acceleration efficiency, but does encode the combined energy contained in accelerated particles and magnetic field. Estimating particle acceleration efficiency from observations is complex, as the population of nonthermal particles in the blob at $90$~days is likely a combination of particles accelerated in the initial radio flare and particles accelerated during the blob's propagation. Thus, neither the particle acceleration efficiency nor mechanism is well known. In the fiducial simulation, further described in section \ref{fidsim},  we adopt $\kappa = 0.1$ which is roughly equivalent to a particle acceleration efficiency of $5\%$. This value is a reasonable choice considering that i) $\kappa$ must be less than unity; and ii) $\sim 10$\% of the shock power is thought to go into cosmic rays in supernova remnants \citep{helder2009,bell013,morlino2013}. Our chosen $\kappa$ is also broadly consistent with results from modelling of radio galaxies \citep{matthews2021} and particle-in-cell simulations of electron-ion shocks \citep{marcowith2020}. 
In reality, the particle acceleration efficiency will depend on a number of factors including $\Gamma$, shock Mach number, magnetisation and magnetic field geometry, and is likely to change over time.

Converting from simulation to luminosity units using equatiom~\ref{eq:emis_to_prs_full}, accounting for the efficiency $\kappa$, is an inversion of the minimum energy calculation. Given the observed luminosity, we implement initial conditions such that the simulation at $t_{\rm sim}=0$ reproduces the measured luminosity at $t_{\rm obs}=90$~days. This process is equivalent to choosing the appropriate minimum energy for our adopted blob radius and accounting for the correct relativistic effects.


The choice of $\kappa$ affects the initial conditions, because
smaller values of $\kappa$ decrease the luminosity, and therefore the initial pressure must be increased to reproduce the fixed initial luminosity of the blob. This modulation of the initial pressure due to the choice of efficiency  has a knock-on effect on the kinematics of the blob. For example, low values of $\kappa$ require high pressure initial conditions and therefore have a significant amount of internal energy -- this results in behaviour that resembles a Sedov-Taylor blast wave (as opposed to a projectile) thus modifying the trajectory to be shallower at late stages. Qualitatively, this is due to the fact that at late times the forward shock propagates into the ISM and separates from the blob material, leaving the radiating component behind to decelerate. Higher values of $\kappa$ seem to better replicate the trajectory, although it is clear that a more robust treatment of particle acceleration and its efficiency is required in order to interpret the luminosity of these simulations. 

We assume that only blob-material is emitting synchrotron radiation, and therefore in the simulation the shocked ISM does not radiate.   We achieve this by weighting the pseudo-emissivity by a passive tracer, initially set to one inside the blob and zero outside. This assumption matches what we see in large-scale AGN jets, where synchrotron emission is seen to outline the morphology of the jet material \citep{carilli2002, hardcastle2020radio}, whereas the bow shocks are rarely seen in radio. It is currently unclear if this assumption also holds for XRBs, where it seems likely that the forward-shock / shocked ISM contributes in some capacity to the radiation observed (see discussion in section~\ref{discuss_shocks}). However, the tracer-weighted pseudo-emissivity seems to replicate the observed deceleration profile better than that without the tracer for the most of the parameter space explored. Furthermore, tracer-weighting ensures a more compact and trailing morphology in the emissivity, as opposed to a propagating bow shock, which roughly aligns with interpretations of observations (for example the ejecta in XTE J155-560 as seen by \citealt{migliori2017evolving}) although we are often limited by $u$,$v$ coverage. The relative role of reverse and forward shocks in XRB ejecta is discussed in more detail by \cite{cooper2025} and \cite{matthews2025}.

\subsection{Relativistic and geometric effects}\label{relativistic_Effects}
The RHD equations are solved in the lab-frame, i.e. an inertial observer frame, and therefore general post-process relativistic corrections to the simulation are not necessary. It is, however, necessary to correct for geometric viewing effects, which affect both the apparent luminosity and timing/deceleration. 

The geometric time delay, due to changing light-travel time as the blob approaches (or recedes from) the observer is
\begin{gather}
    \Delta t_{\rm obs}=\Delta t_{\rm sim}(1\mp\beta \cos{\theta})~,\label{eq:timedelay}
\end{gather}

where $\Delta t_{\rm obs}$ is the elapsed time measured by the observer for the blob to travel a given distance across the sky (moving with some angle to the line of sight to the observer), $\Delta t_{\rm sim}$ is the travel time of the blob between these two points viewed in the plane of the sky (this is the frame of the simulation), $\theta$ is the angle to the line of sight of the observer, $\beta$ is the velocity $v/c$ over $\Delta t_{\rm sim}$, and $\mp$ corresponds to approaching and receding ejecta, respectively. This results in apparent superluminal motion:
\begin{gather}
    \beta_{\rm apparent}=\frac{\beta \sin{\theta}}{1\mp\beta \cos{\theta}}~.\label{eq:superluminal}
\end{gather}

In the simulation, we measure the displacement as a function of time by tracking the centre of synchrotron emissivity per simulation frame. We correct each simulation time-step into the observer frame with Equation \ref{eq:timedelay} where $\beta$ is measured from the displacement of the centre of emission over $\Delta t_{\rm sim}$. The observed displacement is shortened due to projection effects: $d_{\rm obs}=\Delta t_{\rm obs}\beta_{\rm apparent}$.  

This analysis is applied to the entire duration of the simulation, which begins at $t_{\rm sim}=0$ days, equivalent to $t_{\rm obs}=90$ days for the approaching ejection, after which $\Delta t_{\rm obs}\neq \Delta t_{\rm sim}$. Assuming that the receding ejecta of MAXI J1820+070 are identical to the approaching ejecta but launched in the opposite direction, we are also able to reverse the geometric viewing effects on the simulation data to predict a receding ejection. This entails applying the same corrections to the simulations as outlined above with appropriately reversed signs for receding ejections in Equations \ref{eq:timedelay} and \ref{eq:superluminal}. However, the time $t_r$ in the observer frame at which we would expect to see the receding ejection at the same intrinsic comoving age $t_i$ as the approaching ejection seen at observer time $t_a$ depends on the evolution of the Doppler factors as a function of time, where
\begin{gather}
    t_{a,r}=\int^{t_{\rm i}}_{t_{\rm 0}}\delta_{a,r}^{-1}(t')dt'
\end{gather}
and $\delta_{a,r}(t)$ are the Doppler factors of the approaching or receding ejecta, and $t'$ is the comoving time with the ejecta. The Doppler factor is given by
\begin{gather}
    \delta_{a,r} = \Gamma^{-1}(1 \mp \beta \cos\theta)^{-1} ~.\label{eq:dopplerfactor}
\end{gather}
Therefore to calculate $t_{r,0}$ , the time at which we see the receding ejection at the same intrinsic age as the approaching ejection at $t_{a,0}=t_{\rm obs}=90$ days, we assume constant Doppler factor prior to this measurement. This simplifies the relation to
\begin{gather}
    t_{r,0}=t_{\rm a,0}\frac{\delta_{\rm a,0}}{\delta_{\rm r,0}}
\end{gather}
where $\delta_{a,0}$ and $\delta_{r,0}$ are the Doppler factors of the approaching and receding ejecta calculated for constant $\Gamma=2.0$ and angles $\theta=62^{\circ}$ and $\theta=118^{\circ}$ , respectively.

As we are including pseudo-radiation the during post-processing, special relativity is not already incorporated into the derived luminosity, and therefore we have to consider the effects of relativistic beaming. This is done by adjusting the rest frame flux $f_0$ per cell to the observed flux per cell
\begin{gather}
    f = f_0 \delta^{3-\alpha}~,
\end{gather}
where $\alpha$ is the spectral index of the emitting region, and $\delta$ is the Doppler factor of that cell. 

\section{Results}\label{simresults}

\subsection{Fiducial Simulations}\label{fidsim}

We present a fiducial simulation which serves to replicate the deceleration and luminosity as a function of time, within the constraints of the measured initial conditions, with the simplest observation-informed description possible. The initial parameters which achieve this are listed in Table \ref{Tab:table_fid}, where these initial conditions correspond to $t_{\rm obs}=90$ days. Note that densities are expressed as number density assuming an electron proton plasma, but the fluid equations conserve the lab frame mass density of the fluid and are therefore insensitive to the exact composition of the plasma.


\begin{table*}
\centering
\captionof{table}{Initial conditions and derived properties of fiducial simulation. Primed values are in the comoving frame of the blob, unprimed values are in the lab frame. Number densities are calculated assuming a mean particle mass of $m_p$. Initial conditions correspond to $t_{\rm obs}=90$ days. \label{Tab:table_fid}}
\begin{tabular}{lll}
\hline
 {Parameter} & {Value} & Description\\ \hline
 $\Gamma_0$                                & 2.0  & Initial Lorentz factor \\
$R_{\rm blob}$                                & $9.51\times10^{15}$ cm & Blob radius\\ 
$n'_{\rm ism}$& $0.001$ cm$^{-3}$ & ISM number density \\ 
$n'_{\rm blob}$& $0.02$ cm$^{-3}$  & Blob number density \\  
$P_{\rm blob}$                                & $6.81\times10^{-7}$ g/cm\,s$^2$ & Initial blob pressure \\ 
$P_{\rm ism}$                                & $9.82\times10^{-14}$ g/cm\,s$^2$ & ISM pressure \\ 
$\kappa$                                & 0.1               &   Efficiency term such that $j^\prime_{\nu^\prime} \propto (\kappa P)^{(p+5)/4}$         \\ 
\hline
\multicolumn{3}{l}{Derived quantities} \\ 
\hline
$E_{\rm K, blob}$& $2.16\times 10^{44}$ erg & Initial blob kinetic energy, $(\Gamma_0-1) M'_{\rm blob} c^2$ \\ 
$E_{\rm int, blob}$& $2.20\times 10^{43}$ erg & Initial blob internal energy\\ 
$M'_{\rm blob}$&  $2.41\times10^{23}$g~($1.21\times10^{-10}~M_\odot$) & Blob rest mass \\ \hline
\end{tabular}
\end{table*}

The exact choice of initial conditions are intended as a `ballpark' regime for this system, and should not be interpreted as measurements or exact predictions. Instead, we wish to discern general values in order to extract a qualitative description of the behaviour.

The parameters tuned to replicate observational data were primarily the absolute densities of and contrast between the blob and ISM, largely unconstrained by the observations alone. Other free parameters included the efficiency $\kappa$, and the initial Lorentz factor $\Gamma_0$, which were explored within reasonable values, loosely constrained by reasonable assumptions. Discussion of our choice of $\kappa$ can be found in the above section \ref{analysisandunits}.

The Lorentz factor of these ejecta are difficult to constrain from kinematics alone due to relativistic and geometric effects even if their trajectories are well resolved. In \cite{Bright2020}, this is done by fitting the trajectories of the approaching and receding components to a linear model to obtain proper motions, and using the relationship between these proper motions to constrain the product $\beta\cos\theta$, solvable for both $\beta$ and $\theta$ for a given distance. The maximum angle to the line of sight is obtained at $\beta=1$, which also constrains the maximum distance $d_{\rm max}$ to the source with some uncertainty, beyond which a more extreme angle than this maximum is required to explain the proper motions observed. Apart from at very small angles and speeds, the inferred $\Gamma$ from these calculations blows up in the vicinity of $d_{\rm max}$ (where most all relativistic sources are naturally placed due to the asymptotic behaviour of $\beta$ towards 1) and therefore only lower limits on $\Gamma$ are possible \citep{fender2003uses}. \citet{Bright2020} place the lower limit on the bulk Lorentz factor at $\Gamma>1.7$. Additional constraints can be made through the relativistic effects imprinted on the luminosity, but this requires a knowledge of the intrinsic brightness of the system, which is currently unknown. We explore Lorentz factors in the lower range ($1.7\leq \Gamma \leq5$), and the fiducial simulation is run with an initial $\Gamma=2.0$ (and therefore $\theta=62^{\circ}$) as this provided the best visual fit to the data, particularly the curvature of the deceleration curve.

In simulations with higher Lorentz factors (for the same kinetic energy, lower blob mass), the deceleration of the blob occurred much more abruptly than those with lower $\Gamma$, and therefore produced a poorer match to the deceleration at later times. This peculiar behaviour can be attributed to the increase in pressure due to the additional de-boosting of emissivity associated with a higher $\Gamma$, as we calculate the minimum internal energy (and therefore pressure) directly from the inferred intrinsic flux. In high $\Gamma$ simulations, the blob is dominated by internal energy and is quickly obliterated by a strong reverse shock, leaving most of the jet mass behind as the forward shock carries the rest of the energy into the ISM, the same phenomenon seen when increasing pressure to account for a low $\kappa$ values as discussed at the end of section \ref{analysisandunits}. It is therefore difficult for the observed behaviour to be described with significantly higher Lorentz factors for a conserved kinetic energy -- issues with higher kinetic energy are discussed in detail in section \ref{accretion-jet-connection}. 

Lower Lorentz factors also pose problems. Simply lowering $\Gamma$ and therefore also the kinetic energy resulted in too small propagation distances. A proportional increase in the mass of the blob (to conserve kinetic energy) also posed problems due to plausible accretion rates as also discussed in Section \ref{initialconditions}, in addition to the kinematic lower limit at $\Gamma>1.7$.

We explored a wide range of densities for both the blob and ISM, varying both the absolute densities and the contrast. Generally, the higher the density contrast $\chi=\rho_{\rm blob}/\rho_{\rm ISM}$, the further the blob travels before disrupting and decelerating. For the same density contrast, simulations with higher blob absolute density tend to travel slightly further, although the more prominent effect is the increased luminosity at later times due to the larger kinetic energy available for dissipation as thermal energy. It is therefore necessary to balance a high enough density contrast such that the blob remains intact for long enough (and decelerates at late enough times) to match observations, with a low enough absolute density of the blob such that a) the mass does not exceed the back of the envelope limit (see Section \ref{initialconditions}), b) there is no unrealistic significant late-time rebrightening due to energy dissipation of the additional kinetic energy. Lastly, the ISM absolute density was kept within physically reasonable values, i.e. within range of known phases of the ISM derived from \cite{longair1995}.

Tuning these parameters to best match with observations, without fine-tuning, we arrive at the values as listed in Table \ref{Tab:table_fid}. This fiducial simulation requires a low-density ISM akin to a `hot coronal' ISM as described by \citet[][and originally courtesy of Dr John Richer]{longair1995} along with an over-dense blob moving relativistically. 

\subsection{Qualitative Behaviour}

\begin{figure*}
    \centering
    \includegraphics[width=0.7\textwidth]{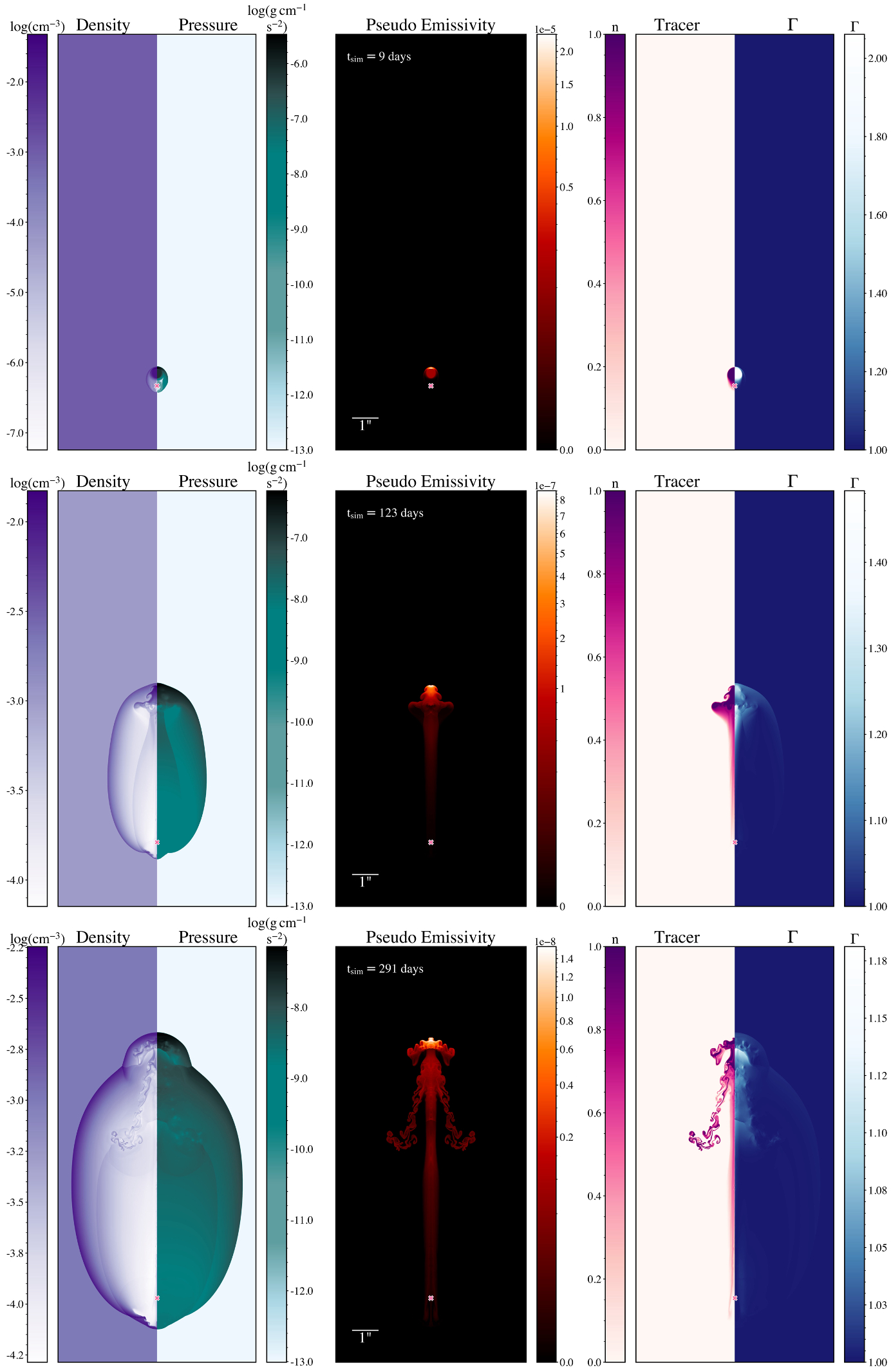}
    \caption{Density, pressure, pseudo-emissivity, tracer, and Lorentz factor of the fiducial simulation (from left to right). From top to bottom, rows are ordered in increasing $t_{\rm sim}$: 9 days, 123 days, and 291 days, where $t_{\rm sim}=t_{\rm obs}-90$. A scale-bar is provided for angular size at 2.96 kpc, while the plot domain is roughly $325\times10^{15}$cm by $650\times10^{15}$cm. The pink and white `X' marks the initial position of the centre of the blob at $t_{\rm sim}=0$ and $t_{\rm obs}=90$. A detailed discussion of the evolution presented here is given in the main text.
}
    \label{fig:multipanel-sim}
\end{figure*}

Figure \ref{fig:multipanel-sim} depicts the evolution of the blob in the fiducial simulation. Each horizontal row of the figure shows a snapshot of the simulation, with time increasing from top to bottom. At each time-stamp, the density, pressure, pseudo-emissivity, tracer, and Lorentz factor are mapped respectively left to right. The far left and far right columns are split down the middle to depict two different parameters on each side -- this is done without information loss thanks to the axisymmetry of the simulation volume. The middle column displays the entire axisymmetric volume to depict what an idealised radio telescope might detect.

In the topmost row, we see the blob very shortly after being launched in the simulation with initial conditions meant to replicate 90 days after its original launch from close to the black hole ($t_{\rm sim}=0$, $t_{\rm obs}=90$). At this point in its evolution, the blob is dense, compact, and moving relativistically. After only 9 days in the simulation, the blob morphology has developed significantly: the tracer has become vertically smeared out along the direction of its trajectory, the emissivity is concentrated in a shock at the head of the blob, and trailing behind the blob a high pressure and low density wake is being created. The maximum Lorentz factor in the blob is slightly greater than the initial $\Gamma$ at launch, likely due to collimation of the blob (we also observe further re-collimation and high $\Gamma$ factors even later in the simulation, see Figure \ref{fig:shock_multitime_line} which we discuss in more detail later).

After propagating for roughly another 3.5 months, as shown in the middle row, the tracer material has become compressed and laterally extended with a long trailing tail, and the high-pressure low-density cavity has continued to inflate. The inner cavity at $t_{\rm sim}=123$ days has a number density  of $\sim2\times10^{-4}$ cm$^{-3}$ and pressure of $\sim 1.0\times10^{-8}$ g cm$^{-1}$s$^{-2}$ , compared to the surrounding medium density $10^{-3}$ cm$^{-3}$ and pressure $9.8\times10^{-14}$ g cm$^{-1}$s$^{-2}$. The cavity is therefore roughly an order of magnitude lower density and five order of magnitude higher pressure than the surrounding medium at this point in the propagation. Certain regions remain significantly relativistic: the blob itself at the head of the jet, as well as a recollimated region behind the blob along the spine of the jet. A powerful forward shock envelops the head of the blob, propagating relativistically into the ISM and powering the expansion of the cavity, while a reverse shock propagates into the blob itself as it decelerates, discussed in more detail in Section \ref{instabilities_shocks}. The pseudo-emissivity continues to be concentrated in the shocked region, specifically tracing the reverse shock.

In the final stages of deceleration, roughly 9.5 months after initial conditions, and almost a year after the likely launch, we see significant development of swirling instabilities in the tracer and density. Much of this rich structure is almost invisible in the synchrotron pseudo-emissivity, due to the lower pressure of these areas, especially in comparison with the shocked region at the head of the blob. As a result, these structures are mostly too faint for current radio telescopes to detect. Exactly what these simulations would look like from the perspective of modern radio instruments is explored in detail in Section \ref{ezblob}. The low-density cavity continues to grow larger in size, with a density of roughly $10^{-4}$cm$^{-3}$ and pressure $2.4\times10^{-9}$ g cm$^{-1}$s$^{-2}$ at $t_{\rm sim}=291$ days.

Throughout its trajectory, the morphology of the blob (both defined by tracer material and synchrotron emissivity) is noticeably non-spherical. In fact, the original spherical morphology at initial conditions is rapidly destroyed even before the significant development of instabilities at later times. It is evident that a spherical morphology is not stable in these conditions, meaning that the blob at $t_{\rm obs}=90$ days was most likely non-spherical as well -- some implications of this are discussed further in Section \ref{sec:model_comparison}.

\subsection{Deceleration profiles} \label{decel_section}

\begin{figure*}
    \centering
    \includegraphics[width=0.7\textwidth]{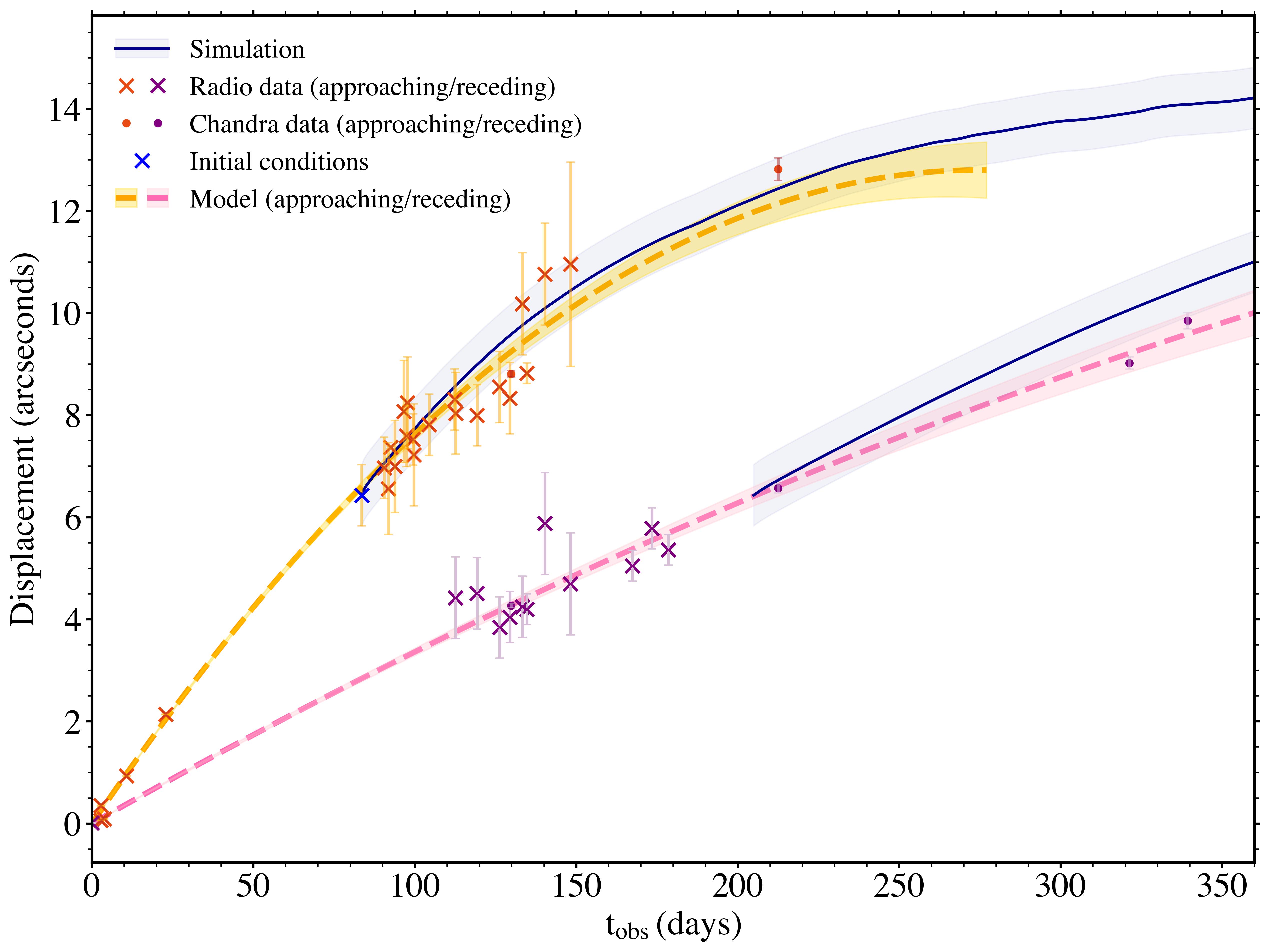}
    \caption{Displacement of the centre of emission of the blob as a function of observer time $t_{\rm obs}$, compared with data. The solid blue line represents the displacement of the simulation centre of emission, with shaded blue error region extrapolating the original error on the displacement measurement. Radio detections of MAXI J1820+070 are shown as crosses, and additional X-ray data of the deceleration from \protect\cite{espinasse2020} are presented as circles. The displacement of the approaching ejection is indicated in orange and receding in purple. The measurement which is used as the initial conditions of the simulation is indicated as a dark blue cross. Deceleration models from \protect\cite{espinasse2020} are dashed lines, approaching ejection modelled in orange/yellow and receding in pink, both with associated errors on the model as shaded regions.}
    \label{fig:deceleration}
\end{figure*}

We measure the displacement as a function of time by tracking the centre of synchrotron emissivity per simulation frame. In order to compare to data, we then adjust for geometric effects due to viewing at an angle to the line of sight using equations \ref{eq:timedelay} and \ref{eq:superluminal}. 

The resulting displacement is presented in Figure \ref{fig:deceleration}. Assuming that the receding ejecta of MAXI J1820+070 are identical to the approaching ejecta but launched in the opposite direction, we are also able to reverse the geometric viewing effects on the simulation data to predict a receding ejection as well, detailed in Section \ref{relativistic_Effects}. 

The displacement as a function of time of the simulation blob for both viewing angles is compared with real data from MAXI J1820+070 as measured in radio by \cite{Bright2020} and X-rays by \cite{espinasse2020}, seen in Figure \ref{fig:deceleration} as orange (radio) and purple (X-ray) crosses. Starting from the conditions at 90 days post-flare, we are able to reproduce the deceleration of both the approaching and receding components qualitatively.

Using a simple constant deceleration model of the apparent proper motion, \cite{espinasse2020} were able to fit both the radio and X-ray data -- this model is also shown in Figure \ref{fig:deceleration} as orange and pink dashed lines for the approaching and receding ejecta respectively.  The simulation is not meant to be a fit to the data, but is being used to qualitatively explain the phenomena behind the deceleration and the regime of the parameters.

By varying simulation $\Gamma$, $\kappa$ and densities of both the ISM and blob, we infer that a blob significantly denser than its surrounding medium is required in order to reproduce the distance travelled. Given the required density contrast, and an upper limit on the absolute density of the blob due to maximum mass and energetics arguments as well as constraints on luminosity of the ejecta as a function of time (discussed in the earlier subsection \ref{fidsim} ), the surrounding medium is required to be (at most) roughly the same density as the hot coronal, low density, ISM phase in order to replicate the trajectory. Figure \ref{fig:deceleration} shows, as a proof of concept, that these relatively simple simulations can, with plausible inital physical conditions, reproduce the deceleration very well. In particular, a low density environment is consistent with the data, suggesting that the ejecta are likely propagating into cavities. 

\subsection{Energetics}\label{energetics}

As XRB ejecta propagate through space and interact with the ISM, significant amounts of energy can be dissipated through various interaction mechanisms. Exactly how energy is transferred between the jet and the ISM is not well understood, although the common framework is through shocks and mixing. In fact our lack of understanding how, where and with what efficiency the initial kinetic energy of XRB ejecta is transferred to the ISM is a major roadblock in trying to extract energy estimates from radio images.

\begin{figure}
    \begin{center}
    \includegraphics[width=\columnwidth]{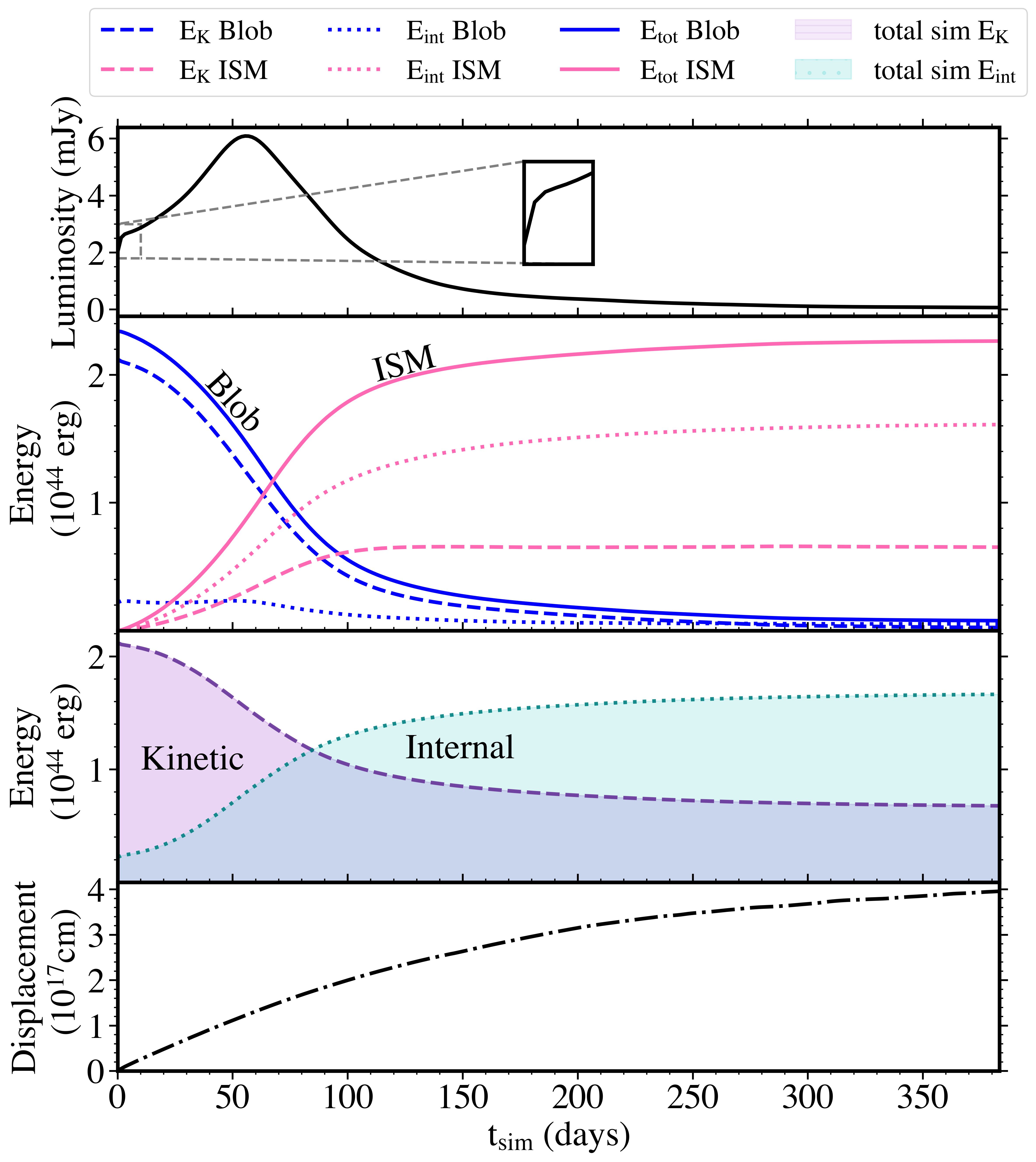}
    \caption{Transfer of internal and kinetic energy in the blob, ISM, and entire simulation as a function of time $t_{\rm sim}$, compared with the luminosity and the deceleration of the centre of emission. The shaded regions refer to energy in the entire simulation volume, whereas the lines refer to energy contained within just the blob or ISM. Solid lines refers to total energy, dotted lines indicate internal energy, and dashed lines refer to kinetic energy. Blob energy is presented as dark blue, and ISM energy in pink. The total simulation kinetic energy is indicated in shaded purple, and total simulation internal energy in shaded turquoise.}
    \label{fig:energetics}
    \end{center}
\end{figure}

Figure \ref{fig:energetics} demonstrates the overall transfer of energy within the simulation, and between the blob and ISM. In addition to this, we also show the observed luminosity as a function of time (scaling with the internal energy in the blob, with additional relativistic effects) and the displacement of the blob as a function of time. 

At early times, most of the energy in the simulation takes the form of blob kinetic energy. In the first few timesteps of the simulation, a portion of the kinetic energy of the blob is transferred into blob internal energy which continues to increase at early times. This behaviour is seen most clearly in the increase in synchrotron flux, producing the lightcurve jump around $t_{\rm sim}\sim0$ days and peak around $t_{\rm sim}\sim65$ days, although this lightcurve is also convolved with relativistic effects and therefore also reflects the changes in $\Gamma$. After this point of maximum flux, the blob loses internal energy and the flux decreases significantly afterwards as this energy is transferred into the ISM. This heating of the blob is mediated by the reverse shock passing through the blob itself. The remainder of the kinetic energy of the blob is transferred directly to the ISM.

The ISM is simultaneously being accelerated and heated by its interaction with the blob, but the energy increase is predominantly through heating. The blob continually accelerates ISM material until about 120 days after initial conditions, after which the kinetic energy in the ISM plateaus until late times. The internal energy in the ISM increases more rapidly compared to the ISM kinetic energy, and tends towards stagnation at later times, between 200-250 days post initial conditions. The blob is therefore more efficient at heating the ISM rather than setting it in motion.

Overall, the simulation depicts a gradual transition from a kinetically to thermally dominated system,
where the turnover occurs around $t_{\rm sim}=80$ days, slightly after the point at which the total energy transferred to the ISM becomes larger than the total energy remaining in the blob. The total energy in the blob is almost entirely lost to the surrounding medium within the simulation time, and this energy persists as roughly $2/5$ kinetic energy and $3/5$ internal energy in the ISM. A minimal fraction of the initial kinetic energy is lost to internal heating, of which another minimal fraction will be radiated away through synchrotron cooling, but inevitably will play almost no role in the dynamics of deceleration. 

The behaviour of the lightcurve is a direct result of the transfer of energy from kinetic to internal within the blob, as well as the choice of particle acceleration efficiency. In this case, our treatment of particle acceleration efficiency is constant in time and across the simulation volume, set with $\kappa=0.1$ in Equation \ref{eq:emis_to_prs_full}. The luminosity is very feasibly powered by this mechanism of energy transfer, but the relationship between internal energy and particle acceleration is likely much more complicated. We demonstrate the possibility of ongoing in-situ particle acceleration, but the significant rebrightening of the blob up to 6~mJy is not a realistic outcome, and therefore requires a more elaborate treatment of particle acceleration efficiency in the simulation. 

\subsection{Disruption of the blob}\label{instabilities_shocks}

\begin{figure*}
    \centering
    \includegraphics[width=1.0\textwidth]{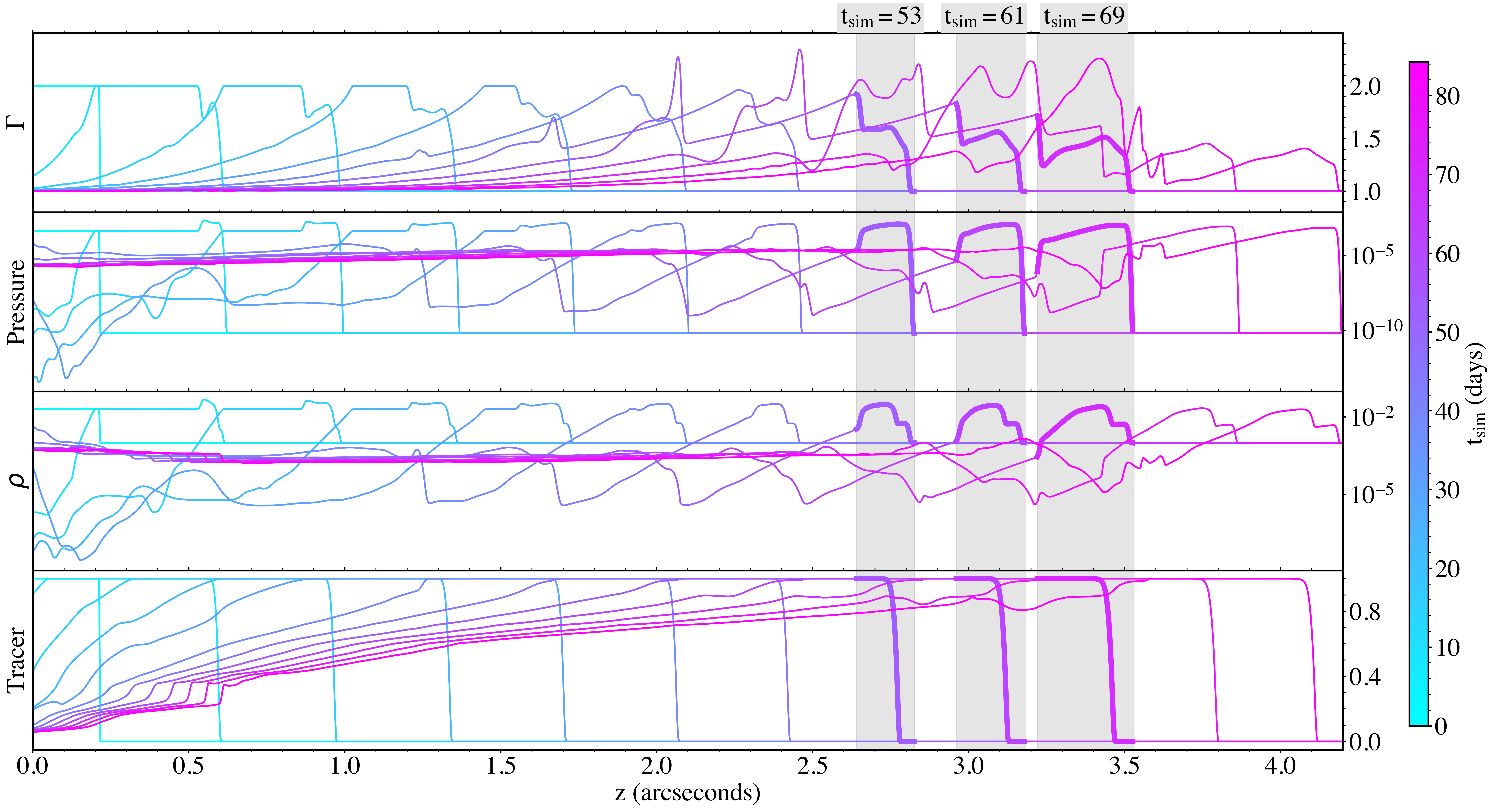}
    \caption{Profiles of simulation variables along the $z=0$ axis (axis of symmetry) as a function of time. The units of the variables are arbitrary simulation units, except for time and distance which are given in days and arcseconds respectively. The initial conditions at $t_{\rm sim}=0$ are shown in the lightest blue line, whereas the $t_{\rm sim}=84$ evolution is the darkest pink line, and each line is separated by roughly 7.7 days in simulation time. The shocked region at the head of the blob is highlighted for three timesteps ($t_{\rm sim}=53,61,69$ days) to illustrate the profile and evolution of the shock. These analyses provide insights into the evolution of shocks in the ejecta, see discussions in Sections \ref{forwardreverse} and \ref{discuss_shocks}.}\label{fig:shock_multitime_line}
\end{figure*}

\begin{figure}
    \centering
    \includegraphics[width=\columnwidth]{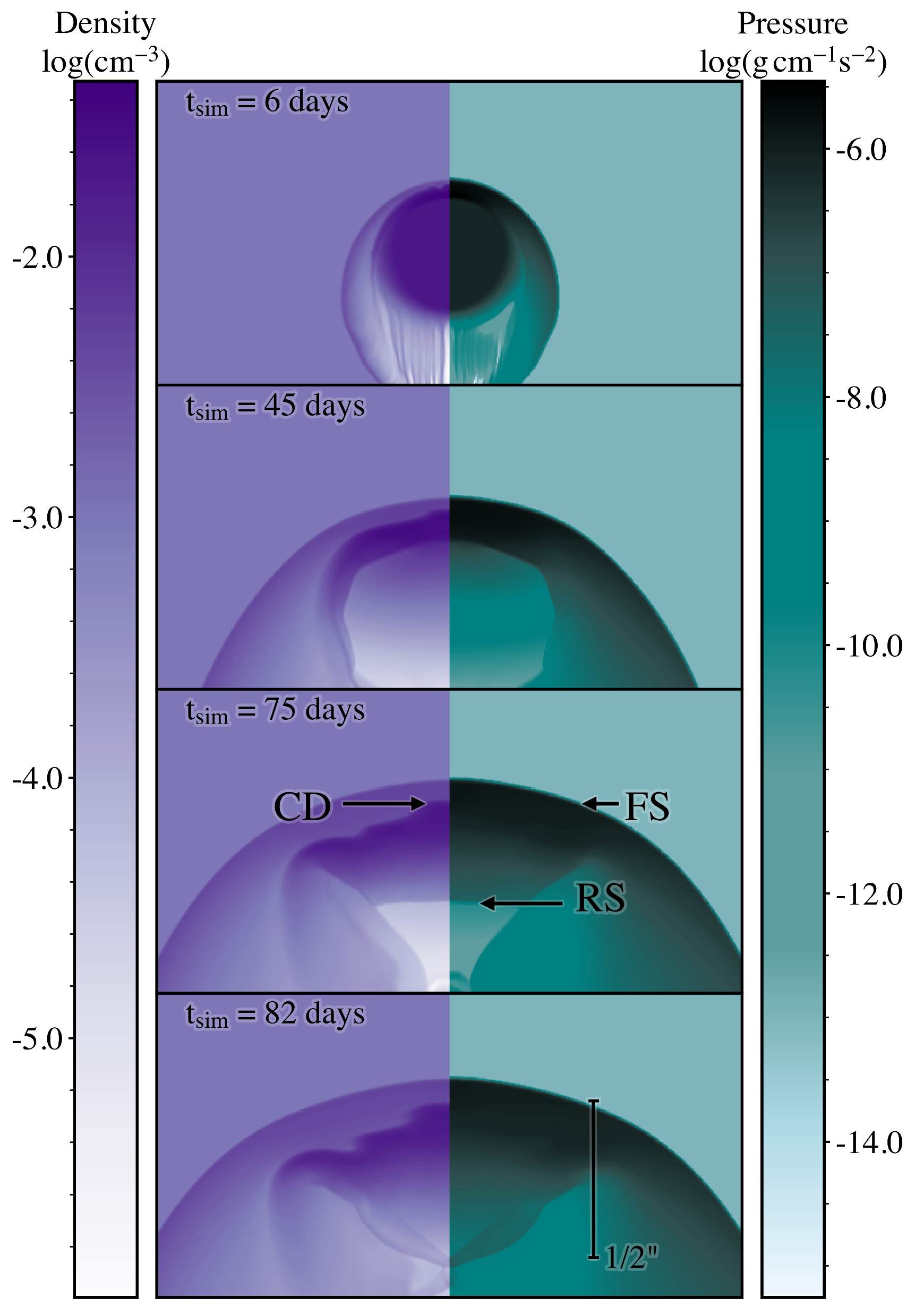}
    \caption{Evolution of the shock structure as a function of time, particularly the reverse shock. Density (left) and pressure (right) at the head of the ejecta are shown for timesteps ranging from $t_{\rm sim}=$ 6 - 82 days, corresponding roughly to the lifetime of the reverse shock. Each panel in the figure is a zoomed inset of the simulation volume centered at the center of emissivity for that frame. In the bottom panel we indicate the 1/2 arcsecond scale as a rough comparison to the size of the blob at initial conditions, assuming 2.96kpc distance. We identify the position of the contact discontinuity (CD), the forward shock (FS), as well as the reverse shock (RS).  }
    \label{fig:reverse-shock}
\end{figure}

\subsubsection{Cloud-crushing}

It is clear from Figure \ref{fig:multipanel-sim} that instabilities (apparent as swirling vortices) and shocks (seen as drastic pressure discontinuities) are produced as the blob propagates. The extent to which these dictate the disruption and deceleration can be understood further by considering the physics of `cloud-crushing', in which a  gas cloud is blown apart by a propagating shock. \cite{klein1994} present a framework for understanding the non-relativistic version of this problem. In this description, the gas cloud is stationary and a shock propagates towards and through the cloud at a given velocity. By shifting into the reference frame of the surrounding medium, this becomes 
essentially equivalent to a non-relativistic plasma blob propagating through a medium and producing a shock. Our situation is thus a relativistic analogue of this well-studied problem.

The characteristic timescale for the disruption of the non-relativistic cloud is the cloud-crushing time, $t_{\rm cc}$, defined as the time for a shock to cross the cloud. This timescale is obtained from ram-pressure balance and is given by
\begin{gather}\label{eq:cloud_crushing_eq}
    t_{\rm cc} = \chi^{1/2}\frac{R_{\rm blob}}{v} ,
\end{gather}
where $v$ is the velocity of the incoming shock wave prior to impacting the cloud, and $\chi$ is the density contrast, given by
\begin{gather}\label{eq:chi}
    \chi = \rho_{\rm blob} / \rho_{\rm ism} ~.
\end{gather}

The cloud dissipates kinetic energy through compression and heating, as well as through dynamical effects. As it travels through the surrounding medium, the cloud is subject to Kelvin-Helmholtz and Rayleigh-Taylor instabilities due to the density contrast and velocity shear at the cloud interface. The instability modes with wavelengths comparable to the size of the blob are the most destructive, and these grow on timescales comparable to $t_{\rm cc}$. It is well motivated theoretically that both of these mechanisms be intrinsically linked to the deceleration of the blob, and thus the cloud-crushing framework provides a useful comparison for understanding decelerating ejecta. Particularly, it makes clear the importance of the high density contrast required for long-term propagation, since $t_{\rm cc} \propto \chi^{1/2}$. Although this is posed in a non-relativistic framework, similar characteristics and timescales are observed in our simulation.

\subsubsection{A forward and reverse shock}\label{forwardreverse}

In addition to the cloud-crushing shock moving backwards though the cloud, there can be a shock which develops at interaction between the surrounding medium and the cloud given a high enough mach number -- the forward/bow shock propagates into the ISM and the reverse/transmitted shock propagates into the cloud. These two shocks are widely recognized phenomena in a host of astrophysical systems including GRBs \citep{gao2015} and AGN \citep{begelman1984,carilli1996}. In GRBs, it is common to observe emission from a short-lived reverse shock and a long-lived forward shock, recently demonstrated in acute detail in the works of \citet{bright2023precise} and \citet{rhodes2024rocking}. Many GRBs are modelled as adiabatically expanding relativistic ``fireballs'' (\citealt{rees1992relativistic}, \citealt{kobayashi1999}) with the characteristic two-shock structure. In this model, the two shocked regions between the forward and reverse shock are in approximate pressure equilibrium with each other and are separated by a contact discontinuity, where the higher density shocked material is propagating into the blob, and the lower density shocked material propagates into the ISM.

Figure \ref{fig:shock_multitime_line} shows the evolution of the simulation parameters along the axis of symmetry in the simulation as a function of time. The initial top-hat function in $\Gamma$, pressure, density and tracer evolves with time along the length of the simulation. In the figure, the shocked region at the head of the blob in several of the later-time snapshots is highlighted with grey panels and bold lines, to make clear the profile of the shock. Within each of these highlighted regions we see the pressure plateau with sharp discontinuities on either side, clearly defining the shocked region. In the pre-shocked blob, the Lorentz factor is high and dramatically decreases as the material upstream of the reverse shock crosses the shock, as is predicted for a relativistic reverse and forward shock structure. Between the shock boundaries, there is also contact discontinuity, which shows up as a jump in density and separates the forward and reverse shock. The forward shock is made up of shocked ISM material, with unshocked ISM in its upstream; the reverse shock is made up of shocked blob material, with unshocked blob material in its upstream. This divide is also made clear by the tracer discontinuity between the forward and reverse shock.

Figure \ref{fig:reverse-shock} depicts this distinct reverse-forward shock structure in 2D. First identifiable at $t_{\rm sim}\sim 10$ days, the reverse shock emerges at early times in the simulation and starts to dissipate after about 80 days, after which the pressure discontinuity at the reverse shock washes out at the back of the blob. The width of the shocked blob material (measured along the direction of propagation) increases as a function of time, and the `washing out' of the reverse shock coincides roughly with this width reaching about the width of the initial blob ($\sim1/2$ arcsecond), and hitting the back of the low pressure low density cavity temporarily formed behind the head of the blob as it ploughs through the ISM. We can interpret this as the crossing time, $t_{\rm cross}$, of the reverse shock, synonymous with the theorized non-relativistic cloud-crushing time. For the parameters in this simulation, $t_{\rm cc}\sim 19$ days, and therefore $t_{\rm cross}\sim 4t_{\rm cc}$. It is therefore consistent that the shock crosses within several cloud-crushing times, as simple non-relativistic assumptions predict. As this simulation begins at $t_{\rm obs}=90$ days, we might expect the reverse shock to cross earlier in the overall evolution.

As well as disruption and consequent deceleration, the reverse shock is also responsible for the majority of the synchrotron emission inferred in this simulation. This is due to the fact that emissivity is weighted by tracer material, and the most strongly shocked material that also contains tracer has passed through the reverse shock. The middle panels in Figure \ref{fig:multipanel-sim} depict how the emissivity is brightest at the head, but the forward shock surrounding the blob is non-radiating. The reverse-shock efficiently thermalises the blob material as it passes through, resulting in the rising lightcurve as seen in the top panel of Figure \ref{fig:energetics}, directly related to the increase in pressure in the blob shown in the panel below. The maximal width of the shocked region, heated by the passage of the reverse shock, coincides with the maximum in the lightcurve. We identify the reverse shock as a candidate site of in-situ particle acceleration, and its propagation responsible for the late-time re-heating of ejecta. In our simulation, this results in significant re-brightening, but with a more refined prescription for particle acceleration could possibly reheat the ejecta just enough to counteract the rapid fading as discussed in \citet{Bright2020}. 

\subsubsection{Instabilities}

\begin{figure*}
    \begin{center}
    \includegraphics[width=\textwidth]{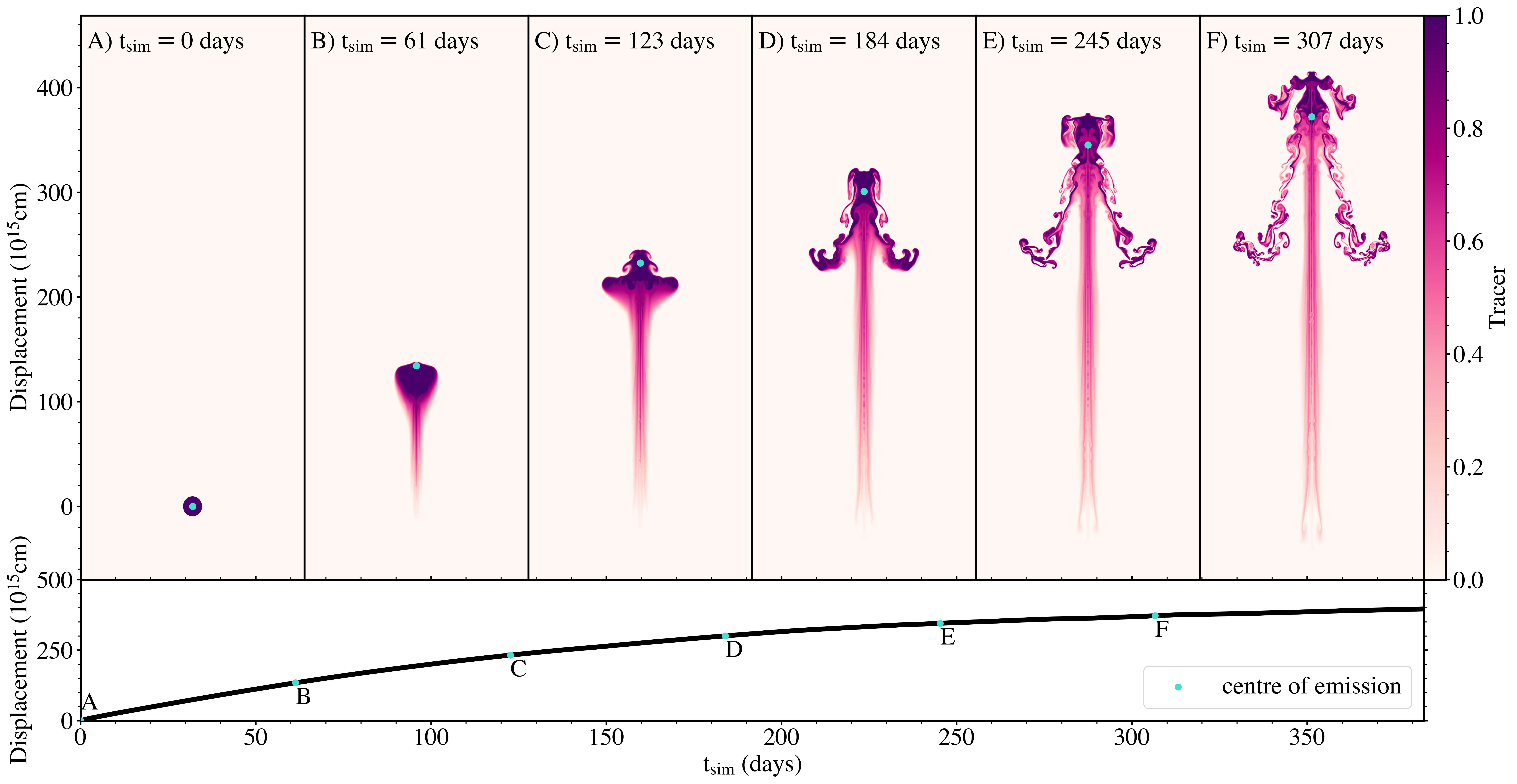}
    \caption{Evolution of the tracer as a function of time. The first row of panels depicts the tracer at various timesteps in purple, and the light blue dot signifies the location of the center of emission in that panel. In the lower panel, the displacement of the center of emission as a function of time is shown in black, and the blue dots indicate the point in time at which the above panels are sliced from.}
    \label{fig:disruption}
    \end{center}
\end{figure*}

At least up until the crossing of the reverse shock most of the energy transfer has been driven by shocks, as significant instabilities have not begun to develop. After this initial shock-crossing phase, coincident with $\sim 4t_{\rm cc}$, instabilities develop significantly enough to begin disrupting the blob. This is the expected timescale for the growth of Rayleigh-Taylor and Kelvin-Helmholtz modes with wavelengths $\sim R_{\rm blob}$ \citep{klein1994}. Instabilities are most noticeable in the tracer fluid, and in Figure \ref{fig:disruption} we show the blob tracer as a function of time, as well as the position of the center of emission in each frame. We see that instabilities growing on scales similar to the size of the blob develop between panels B) and C), just after the reverse shock crossing. These instabilities, in particular the Rayleigh-Taylor instabilities producing `umbrella-like' filaments protruding at the head of the blob, and Kelvin-Helmholtz instabilities swirling at the velocity shear layer along the side of the blob, significantly impact the blob morphology at late times (although we note that these likely develop earlier in reality, as our initial conditions are set at $t_{\rm obs}=90$ days). The large turbulent eddies produced become visible in the emissivity in Figure \ref{fig:multipanel-sim}, and thus could feasibly be detected by high resolution observations, discussed further in section \ref{ezblob}. It is clear that the instabilities must also affect the dynamics by dissipating energy in conjunction with the forward shock, but the extent to which they contribute to the overall deceleration of the blob is not immediately obvious. 

Considering that the rate of change of the ISM internal energy is greater than that of the ISM kinetic energy until late times, and assuming that the main mechanism through which instabilities dissipate energy is kinetically (we do not observe significant pressure increase in zones of instabilities), one can infer that instabilities must play a secondary role in deceleration as compared to shocks. In addition to this, the rate of deceleration varies quite smoothly and is not significantly impacted by the onset of large instabilities.

To more formally assess the relative role of shocks and instabilities in determining the deceleration and energy transfer, we ran a one-dimensional (1D) analogue simulation. The 1D simulation was run with a spherical geometry with initial conditions identical to those set for the fiducial simulation, except as a 1D expanding shell. In this geometry, instabilities are inherently suppressed, and it therefore isolates the effects of shocks as deceleration mechanisms. The 1D shell followed a very similar trajectory as compared to the fiducial simulation, but propagated $\sim10\%$ further at 375 days, and began to deviate from the fiducial trajectory after about $\sim150$ days after which the fiducial simulation flattens out. This is consistent with the description of instabilities only playing a role at later times (post shock-crossing), and only contributing in a subdominant way. We note that the relative importance of instabilities is likely dependent on the parameters of the simulation, such as $\chi$ and $\Gamma$, as well as initial energy.

\begin{figure}
    \centering
    \includegraphics[width=\columnwidth]{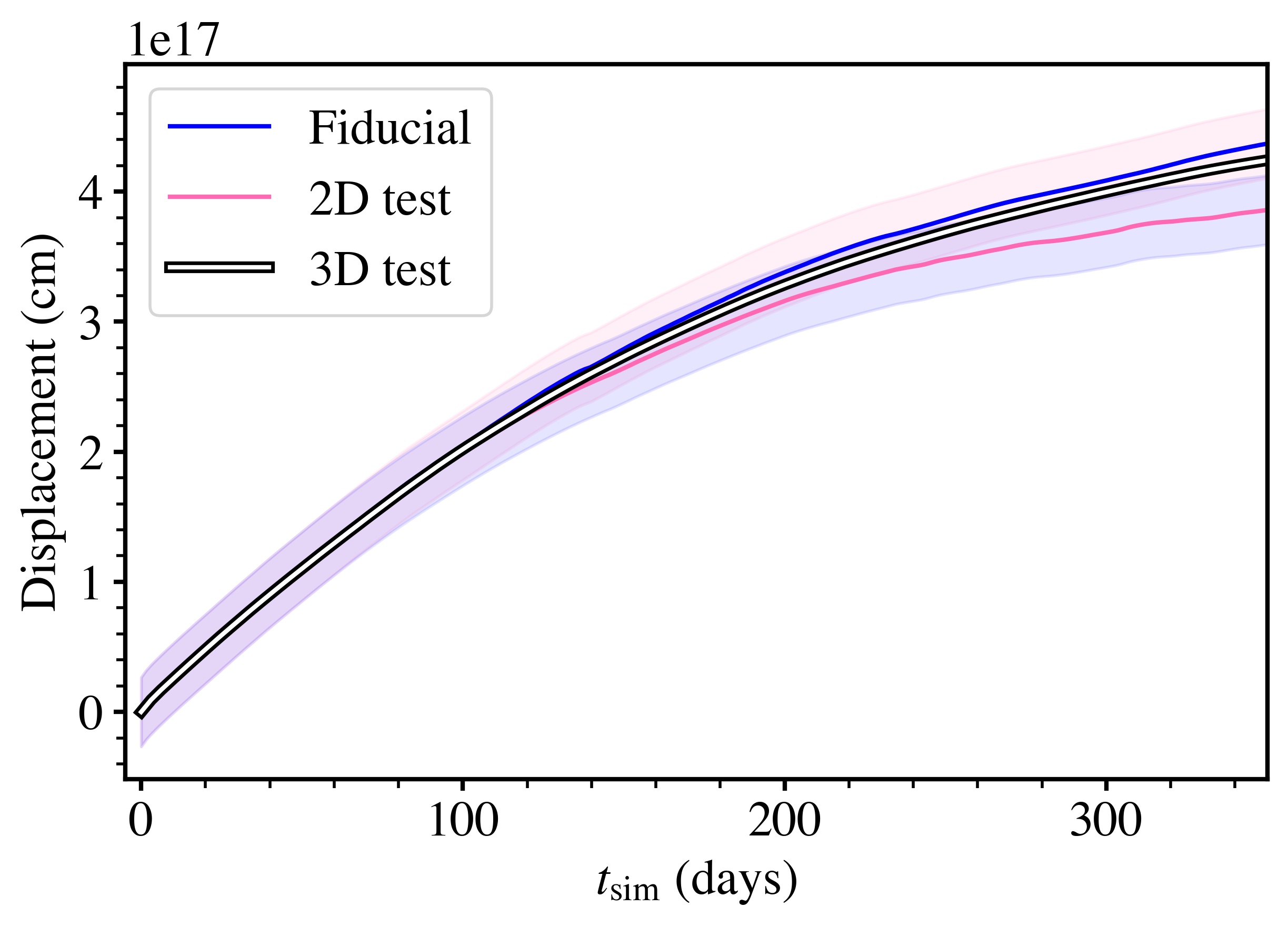}
    \caption{Displacement as a function of time for the 3D simulation ({\sl 3D test}), compared to the fiducial 2D simulation and a 2D simulation with the same resolution as the 3D simulation ({\sl 2D test}). The shaded regions show the (observational) error on initial position from Fig.~\ref{fig:deceleration}. There are some small differences in trajectory for both different resolutions and different dimensionality; however, in each case, the difference in propagation is within the errors and the shape of the curve is very similar.}
    \label{fig:3d_trajectory}
\end{figure}

\begin{figure}
    \centering
    \includegraphics[width=\columnwidth]{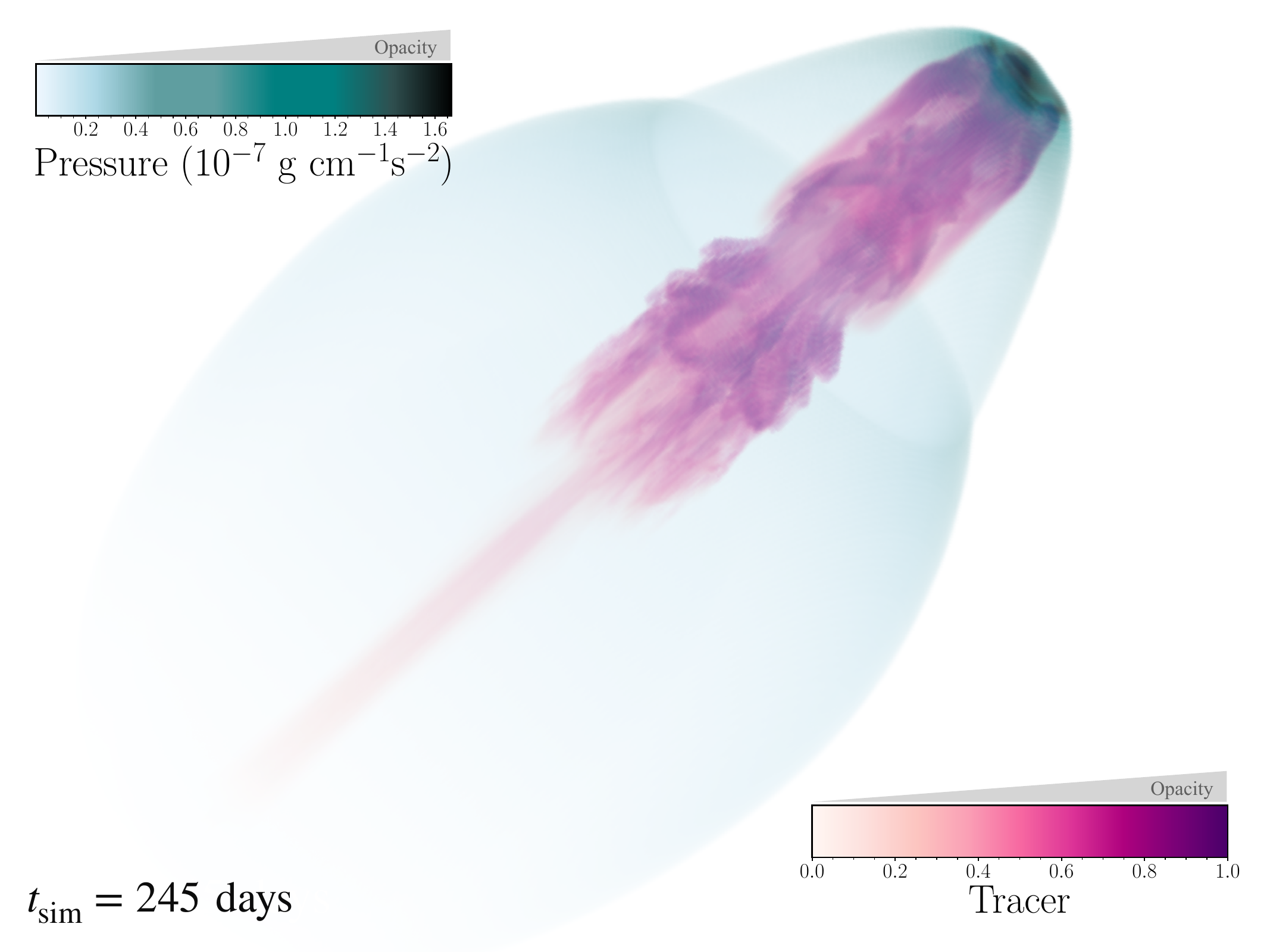}
    \caption{A frame at $t_{\rm sim}=245$ days showing a volume rendering of the 3D Cartesian simulation. The pink indicates tracer, and the teal indicates pressure. Gaussian perturbations have been added in both the ISM and blob density in order to introduce asymmetries.}
    \label{fig:3dsim}
\end{figure}

\subsection{3D simulations}\label{3dsims}

We ran our fiducial simulation in 2D cylindrical coordinates. The 2D simulation is far less computationally expensive than 3D, but it is essential that it displays similar physical behaviour as a 3D simulation. We ran quasi-identical simulations in 3D and 2D at a slightly lower resolution of $1000 \times 1000 \times 2000$ cartesian cells, domain size of $500 \times 500 \times 1000$ simulation units. Apart from geometry and resolution, the only difference compared to the 2D simulation was the injection of Gaussian density perturbations into both the blob and ISM of the order $\delta\rho/\rho \approx 10^{-3}$ in order to break the azimuthal symmetry.

In Fig.~\ref{fig:3d_trajectory}, we show the displacement as a function of time for the 3D simulation, compared to our fiducial simulation and 
the matched resolution 2D simulation. 
The 3D simulation produces a very similar deceleration profile, demonstrating that the physics governing the parameters we are interested in constraining is broadly preserved in 2D, and 3D effects alone do not change our conclusions about deceleration timescales, energetics, or the ability of our simulation to match the observed blob propagation. 

Despite this useful equivalence in a lower dimension, 3D simulations allow us to explore 3D structures both in the blob and in the surrounding ISM. For example, in high-mass XRBs there are strong stellar winds produced by the companion star that interact with the jet, and the clumpy density profile of the wind can highly impact the dynamical and radiative evolution of the jet \citep{perucho2008interaction, perucho2012simulations}. We do not explore these effects here, but we can look for any changes in, for example, blob morphology or the characteristics of the turbulence. Fig. \ref{fig:3dsim} shows a volume rendering of the decelerated blob at $t_{\rm sim}=245$ days. As in 2D, the blob breaks up due to instabilities, with a forward shock propagating ahead of it into the ISM. We find that the physics governing the propagation of the blob and the observed signatures are broadly unchanged, which is unsurprising given the agreement in Fig.~\ref{fig:3d_trajectory}. A reverse shock still passes through the blob, instabilities develop at late times, and a forward shock (visible in teal) propagates into the surrounding medium. However, there is, as expected, 3D structure to the turbulence, rather than an azimuthally symmetric morphology.

\section{Synthetic radio images}\label{ezblob}




Comparing simulation results to observational data can be challenging in radio astronomy, where interferometric images of the same source can vary drastically depending on the array and observing configuration. For this reason, we are developing a pipeline to convert the output of hydrodynamic simulations into simulated radio images which takes into account the exact $u$,$v$ coverage of the proposed, or existing, observation. This pipeline, which we name \texttt{BLOB-RENDER}, we use in its preliminary version to convert the results of the fiducial simulation in this work into pseudo-radio images.

We produce these images with two goals in mind: first, to be able to validate the initial conditions of our simulation by comparing them directly to the real MeerKAT and eMERLIN observations they are informed by; and secondly, to make predictions of what a next generation high-resolution radio telescope might have observed at late times in the deceleration. We will first provide a brief description of how the \texttt{BLOB-RENDER} pipeline works, and then present the results of both of these tests.

\subsection{\texttt{BLOB-RENDER}}

The input data to \texttt{BLOB-RENDER} are the emissivity data from the fiducial simulation. The grid in the simulation is 2D cylindrical, so in order to create an image of the 3D structure, the 2D output is rotated $360^\circ$ and interpolated onto a 3D Cartesian grid, and then integrated along the sight-line into a 2D Cartesian image. This method assumes that the source is optically thin, which \citet{Bright2020} confirm from the spectrum of the radio source, and is generally the case for transient XRB jets. The code-based units for pseudo-emissivity are converted to Jy/pixel according to Equation \ref{eq:emis_to_prs_full}, producing the basic input image for the \texttt{BLOB-RENDER}. 

We then use routines from the \texttt{WSClean} \citep{offringa-wsclean-2014} package, a flexible imaging package for interferometric data, to predict a set of simulated visibilities from the simulation image data. We wish to compare the pseudo-images to real data, so we know the telescope configuration, ie. the exact $u$,$v$ coverage of the telescope, at the time the data were taken and are therefore able to predict visibilities for comparison with this specific dataset. We use the pre-existing MeasurementSets (MS) containing the MeerKAT and eMERLIN observations from \citet{Bright2020} at $t_{\rm obs}\sim90$ days, and thus the correct baseline and baseline projection information. The Jy/pixel data are packaged into an appropriate \texttt{FITS} file format, and supplied to \texttt{WSClean} as a model with which to predict visibilities given the same $u$,$v$ tracks as in the MS. Next, we use \wsclean{} to image these resulting visibilities, producing a predicted radio image of the model.

\subsection{$t_{\rm sim}=0$ radio images}\label{t0_blobrender}

\begin{figure}
    \centering
    \includegraphics[width=1\columnwidth]{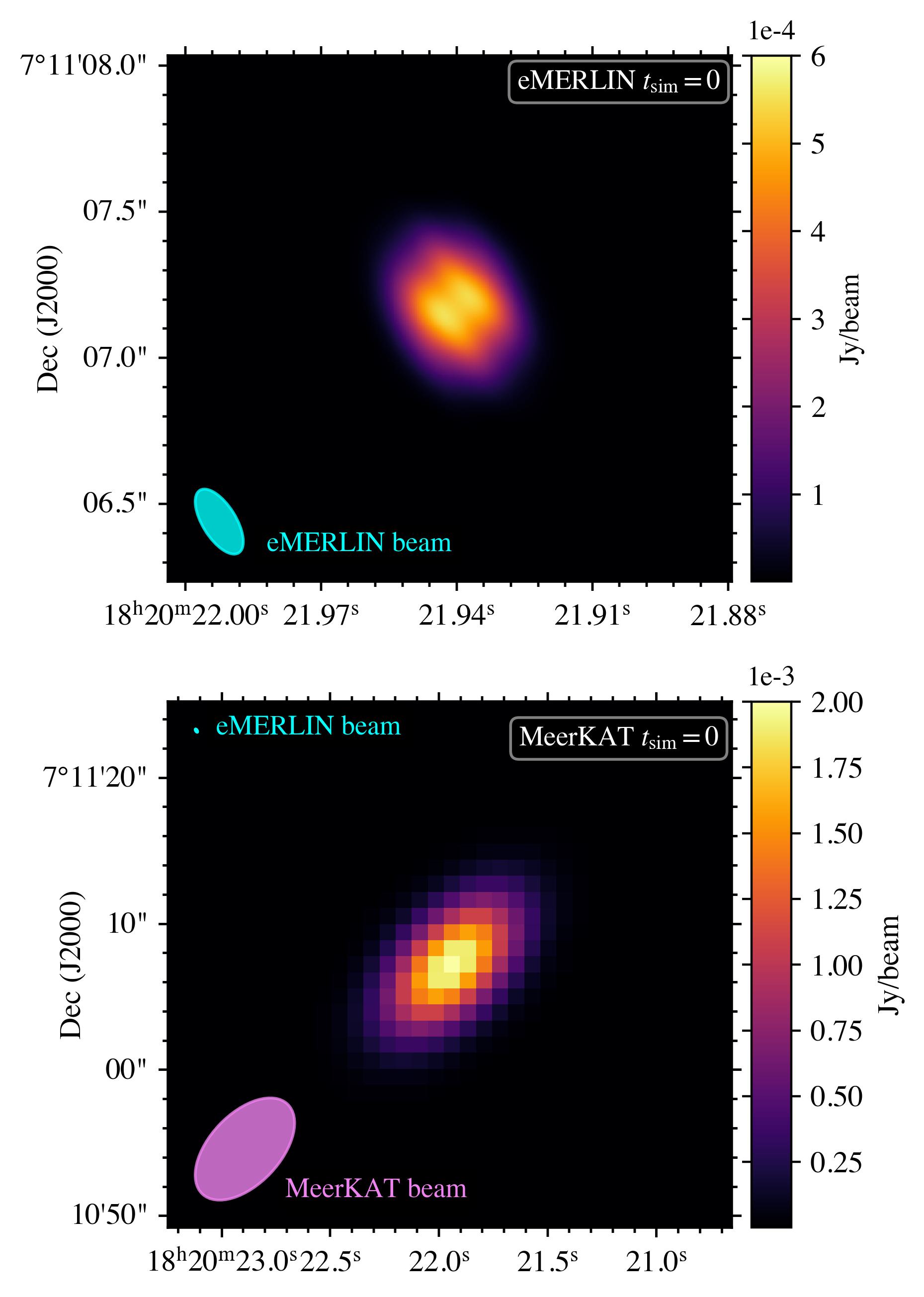}
    \caption{\texttt{BLOB-RENDER} output of simulation initial conditions ($t_{\rm sim}=0$ days) as seen by the eMERLIN (top panel) and MeerKAT (bottom panel) radio telescopes. Both images have no injected noise, but are simulated at the position of the MAXI J1820 core, with the same configurations as the original observations. The respective synthesized beam sizes are indicated in the bottom left of each panel, and the eMERLIN beam is included in the MeerKAT image as a size comparison.}
    \label{fig:blob-render-t0nonoise}
\end{figure}

\begin{figure*}
    \centering
    \includegraphics[width=0.9\textwidth]{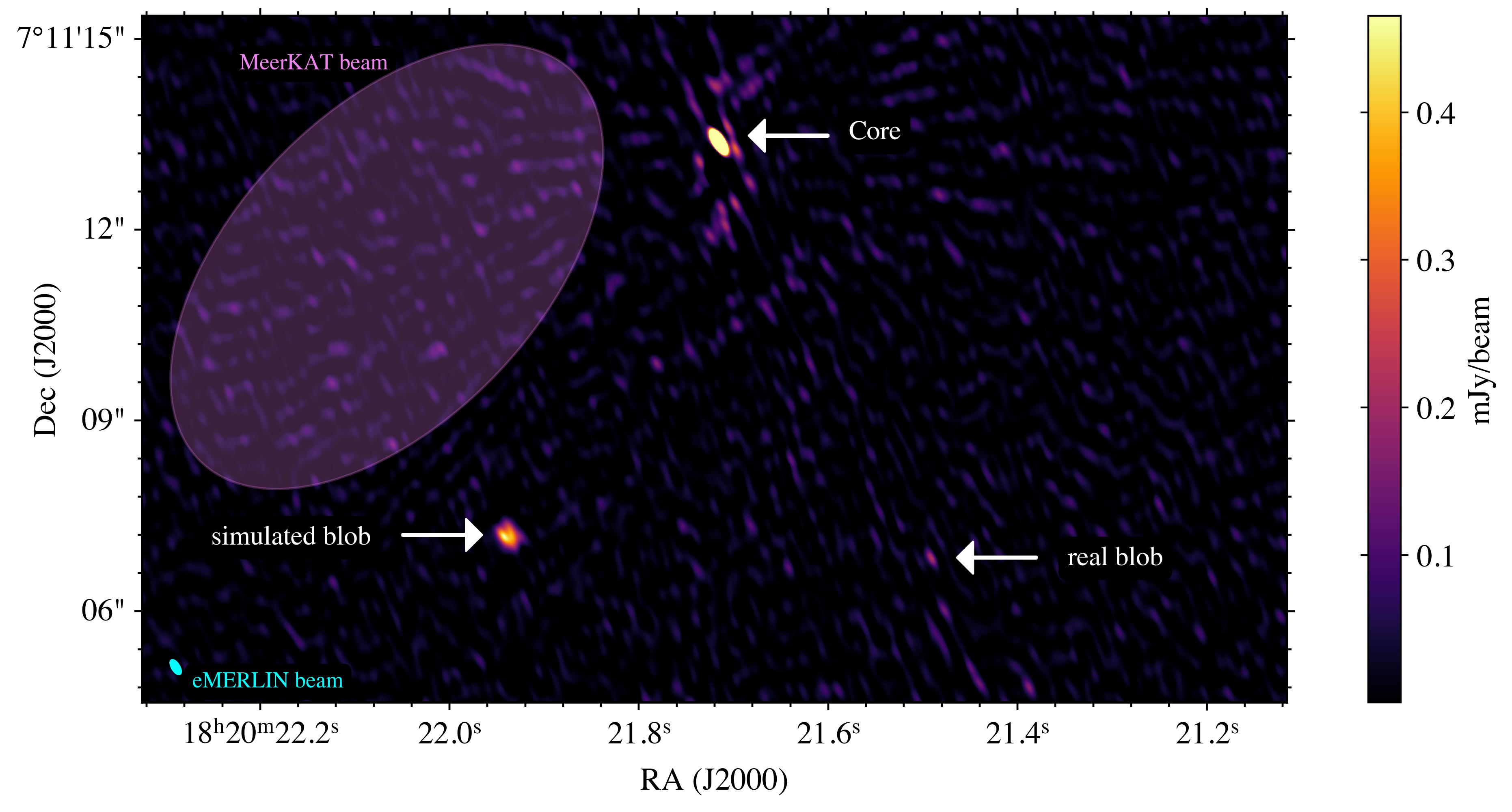}
    \caption{The 1.51GHz radio image of MAXI J1820 at $t_{\rm obs}=90$ days as seen with eMERLIN by \citet{Bright2020}, with the core and ejection (real blob) indicated. We inject the simulated blob at initial conditions ($t_{\rm obs}=90$ days, $t_{\rm sim}=0$ days) into the bottom left of the image to indicate what the effect of background noise is on the structure of the simulated blob. The eMERLIN synthesized beam is indicated in the bottom left, with the MeerKAT beam also indicated in the top left for size comparison.}
    \label{fig:blob-render-real}
\end{figure*}

The two observations at $t_{\rm obs}\sim90$ days constrain the initial conditions of the simulation, and so we predict both the MeerKAT and eMERLIN images of the $t_{\rm sim}=0$ simulation emissivity as a verification step, presented in Figure \ref{fig:blob-render-t0nonoise}. The MeerKAT prediction reproduces the real observations, an unresolved source, with an integrated flux of 2.0 mJy. 

The eMERLIN prediction presents a slightly more complicated picture, and does not exactly reproduce the real data. In reality, eMERLIN quasi-simultaneously detected 85\% less flux than MeerKAT in the same frequency band, but the blob remained unresolved in the eMERLIN image. This was interpreted in \citet{Bright2020} as the majority of the flux being resolved out, therefore lying on scales larger than those probed by eMERLIN. Thus, a size constraint was derived as larger than the eMERLIN synthesized beam but smaller than the MeerKAT synthesized beam. In the fiducial simulation, we choose an initial size of the blob within this range, closer to the lower end of the range, and implement a pressure in the blob to reconstruct the minimum energy (and therefore total flux) for that size -- that total flux is recovered in the simulated MeerKAT image. However, the eMERLIN simulated image also recovers almost all of the total flux, $\sim1.9$m~Jy, and the blob is resolved in the image. The morphology of the blob has been altered by the response of the eMERLIN telescope, and as seen in the upper panel of Figure \ref{fig:blob-render-t0nonoise} this results in a double-peaked resolved structure.

The size constraints derived in \citet{Bright2020} are sufficient for basic minimum energy calculations, but are not accurate size constraints for reconstructing an image. This is due to several simplifying assumptions. Firstly, the constraints are bookended by the sizes of the synthesized beams of the telescopes, but structures larger than the eMERLIN synthesized beam can be imaged thanks to shorter baselines in the array -- in fact, eMERLIN is sensitive to structures up to $2\arcsec$ across despite a synthesized beam of only $0.2\arcsec\times0.31\arcsec$. This means that if in fact the `missing flux' between the eMERLIN and MeerKAT observations was resolved out by eMERLIN, this flux would have to lie on much larger angular scales than previously predicted. Essentially, the morphology of the blob would have to be such that 15\% of the flux density would be from a compact core component, no larger than $0.2\arcsec$ across, and a much larger structure, with no resolvable small-scale structure on scales between $0.2-2\arcsec$ but smaller than $5.4\arcsec$, makes up the rest of the flux. The large separation in size scales between the proposed compact and extended structure is surprising if true but not impossible. 

Secondly, since it is the Fourier components of the images that are being detected, the flux density recovered in the image plane depends on the particular morphology of the blob. We can rule out particular morphologies, but there is no unique solution to the shape of the blob given incomplete $u$,$v$ coverage. We can therefore only speculate possible shapes of our source. 

Lastly, it is not necessary that `missing flux' must be resolved out by eMERLIN --  it is possible that this flux is indeed on angular scales resolvable by eMERLIN but below the sensitivity threshold of the telescope. To see the effect of noise on the simulated image, we have added the predicted eMERLIN visibilities to the original observation data, seen in  Figure \ref{fig:blob-render-real}. We inject the data offset to the core by re-phasing the real data to a new phase centre (an area of the image dominated by noise) and then add the simulated visibilities to these re-phased real visibilities to produce the composite image. The simulated blob is bright and resolved, and the noise has a minimal effect on the overall recovered flux from this image, mainly due to the compact morphology of the source. 

If the flux was concentrated in the centre surrounded by diffuse emission such that the extended emission were below the detection threshold of eMERLIN, this could reproduce the observed data. For example, the RMS noise in the real data is roughly $50\mu$Jy beam$^{-1}$ -- to hide the $1.7$mJy of missing flux beneath the noise it would have to be spread over a minimum of $6.6$ sq$\arcsec$, corresponding to a circular region of radius $\sim1.5\arcsec$. With this flux density morphology, the minimum angular size is still larger than the synthesized beam (the original estimate) but avoids the difficulty of configuring a morphology where 85\% of the flux is on scales larger than $2\arcsec$ and essentially no flux is between $0.2-2\arcsec$, and still recovers the correct flux density with MeerKAT. We note that this somewhat curious structure, a compact `core' surrounded by diffuse emission, is not entirely unlike what is seen at late times in the simulation. Our initial conditions (spherical blob) are of course unrealistic but rapidly evolve to more plausible morphologies.

\subsection{Late-time evolution radio images}

\begin{figure}
    \centering
    \includegraphics[width=0.99\columnwidth]{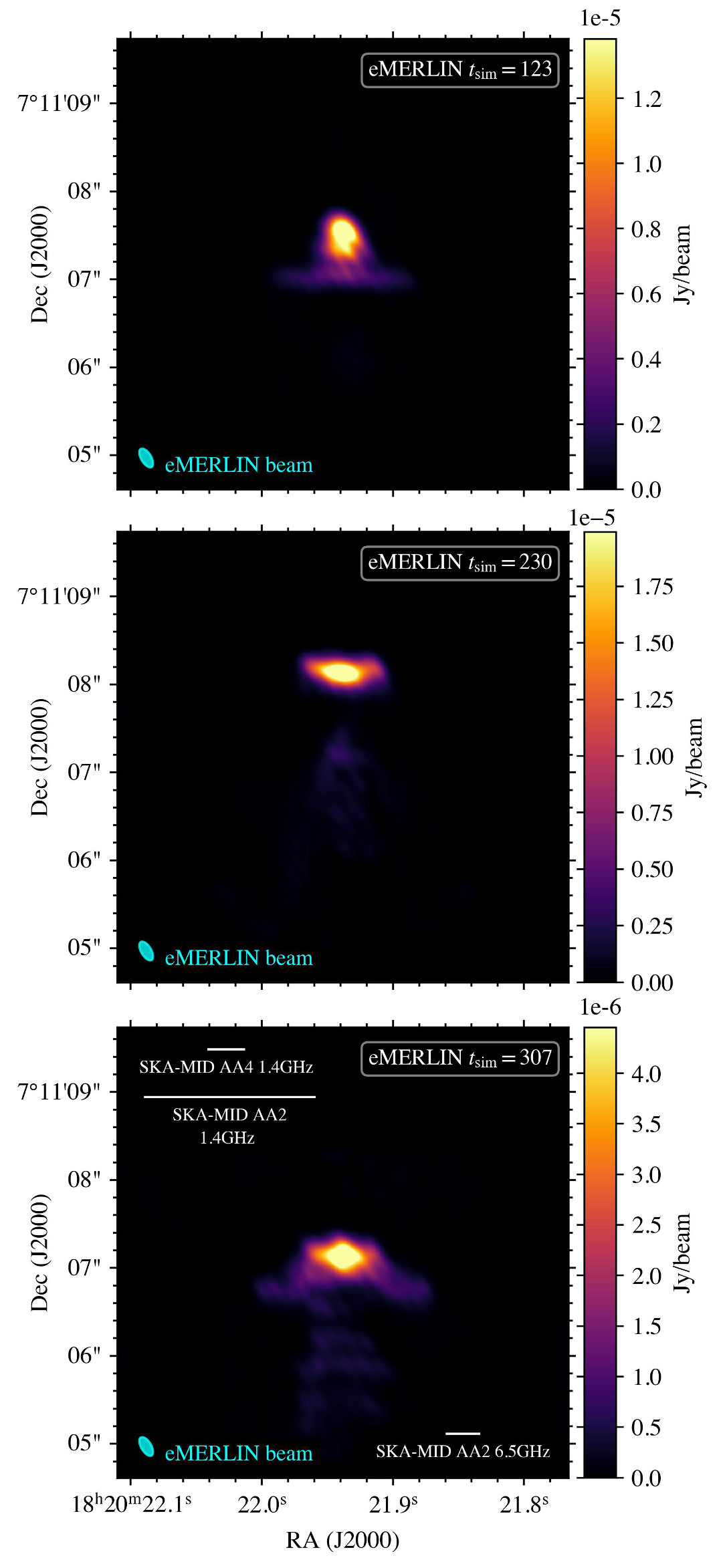}
    \caption{\texttt{BLOB-RENDER} prediction of the fiducial simulation at three timesteps as seen by eMERLIN with identical telescope configuration as the MAXI J1820 observations (where each blob is placed roughly at the phase-centre of the observation) No noise is included in these images. The eMERLIN beam is included in light blue, and the estimated angular resolution of two SKA-MID baseline configurations are included for reference.}
    \label{fig:blob-render-latetime}
\end{figure}

After the spherical initial conditions, the morphology evolves rapidly. At early times, prior to the crossing of the reverse shock and therefore the onset of large instabilities, the emissivity is concentrated at the head of the blob and resembles a laterally extended bow-shock. However, as instabilities begin to disrupt the blob, complex turbulent structures, as well as shocks, are formed downstream. These structure can be seen in simulated eMERLIN images in Figure \ref{fig:blob-render-latetime} -- note that these do not include noise as the RMS noise exceeds the maximum flux density of the simulated images and they would therefore be below the sensitivity threshold of eMERLIN and go undetected. This is expected, as the subsequent real eMERLIN images were also only upper limits. These noise-free images instead demonstrate what kind of morphology we might expect from a blob 80+ days after launch. The flux distribution seen in Figure \ref{fig:blob-render-latetime}, especially at $t_{\rm sim}=307$ , corroborate the hypothesis that the morphology at 90 days post-launch could be described as a compact source with a relatively high flux surrounded by dimmer diffuse emission, where both components are resolvable by eMERLIN but the diffuse emission could lie below sensitivity limits. 

Some of the more detailed structure in these images is also of particular interest. At $t_{\rm sim} = 230$ and $t_{\rm sim}=307$ in Figure \ref{fig:blob-render-latetime} especially, we observe some lateral extension of the blob, as well as some peaks in emission along the direction of propagation. This is compelling, as XRB discrete ejecta have been seen to `split', and even move backwards, as they propagate (see observations of J1727.8-1613 in Hughes et al. in prep and Carotenuto et al. in prep, as part of the ThunderKAT/X-KAT program), as well as extend perpendicular to the jet direction (in the NS XRB Scorpius X-1 \citep{fomalont2001scorpius}, and BH XRBs XTE
J1752-223 \citep{yang2010deceleration} and XTE J1908+094 \citep{rushton2017resolved}). It is possible that `splitting' ejecta could indeed just be multiple ejecta launched in succession, as seen before in GRS 1915 \citep{fender1999merlin} for example, but these \texttt{BLOB-RENDER} simulations demonstrate that even singular launches could result in multi-component images, and unexpected structures due to downstream re-brightening and incomplete $u$,$v$ coverage.

Although these structures are interesting, we can not yet make \emph{direct} comparisons to real images. This is because the striation seen in the 3D projections is to some extent a result of the rotation of 2D cylindrical simulations about their axis. Nevertheless, these images demonstrate how resolving these ejecta at late times may give us insight into the physics of jet deceleration, blob disruption, late-time re-brightening and possible particle acceleration, which we plan to explore in future work. In particular, the future Square Kilometre Array mid-frequency instrument (SKA-Mid), the extension of the current MeerKAT telescope, may be the most likely candidate (along with ngVLA) for being able to resolve these detailed structures. The final design baselines\footnote{\url{https://www.skao.int/en/science-users/ska-tools/494/ska-staged-delivery-array-assemblies-and-subarrays}} (AA4) of the SKA-MID extend up to 159km, providing angular resolution down to $\sim 0.3\arcsec$ at 1.4GHz, comparable to the angular resolution of eMERLIN but with 1) more complete $u$,$v$ coverage and 2) higher sensitivity. The currently funded baseline configuration (AA2), however, only extends to 36km providing a $\sim1.3\arcsec$ angular resolution at this frequency, which may barely resolve these sources at late times (although worth noting that these simulations correspond to the lower end of the size constraint, and also that this is the absolute maximum resolution, and will require a major sensitivity sacrifice to achieve since the array is so core-dense). Both of these scales can be seen with respect to the size of the blobs in Figure \ref{fig:blob-render-latetime}. The higher frequency receivers will provide much better angular resolution, down to $\sim0.3\arcsec$ at 6.5GHz in the AA2 design. However, only a fractions of the dishes will have these capabilities at the AA2 stage, limiting the sensitivity of the telescope at higher frequency, where the source will appear intrinsically dimmer due to the optically thin emission. This overall demonstrates a science case for the inclusion of the longer baselines in SKA-MID.

\section{Discussion}\label{discussion}

\subsection{The role of shocks: acceleration and energetics}
\label{discuss_shocks}
\subsubsection{The reverse shock as a particle accelerator}

It is well established that shocks, in conjunction with turbulent magnetic fields, can be responsible for significant particle acceleration. In particular, late-time re-brightening of transient ejecta from BHXRBs can be caused by in-situ particle acceleration due to shocks as the blob interacts with the surrounding ISM.  
In our simulated blob, shocks dissipate the initial kinetic energy into the ISM and partially into the blob itself. The reverse shock passing through the blob, compressing and heating it, presents a likely site for this particle acceleration and can be responsible for the observed re-brightening. The reverse shock thus establishes a connection between observed deceleration and observed radiation (although we caution the reader that observations are also affected by changing Doppler factor, and a decelerating off-axis jet will naturally become brighter in the absence of any internal evolution). 

\cite{espinasse2020} find that the spectra of the ejecta from MAXI J1820+070 at 130 days post-flare are consistent with synchrotron radiation from particles accelerated up to 10 TeV, likely produced in shocks, consistent with results found for the large-scale jets of XTE J1550-564 \citep{corbel2002large}. A reverse shock, such as the one in this simulation, is a strong candidate for the production of these high-energy particles. The characteristic maximum energy achievable through particle acceleration is the \cite{hillas1984} energy, given by $E_{H}\sim (Z_{e}v_s B R)$. For our adopted values of $\kappa$ and $\eta$ we estimate a magnetic field strength of $B \sim 1~{\rm mG}$ based on the approximate post-shock pressure at 75 days (in the simulation, at which point the reverse shock is about halfway through the blob) of $P \approx 6\times10^{-8}$ g cm$^{-1}$ s$^{-2}$. For $v_s = c$ and a transverse shock width of roughly $10^{16}~{\rm cm}$ at 75 days, this gives a maximum energy estimate of $E_{H} \sim 3~{\rm PeV}$. This is a reasonable estimate for protons in ideal particle acceleration conditions; if we instead calculate the electron maximum energy as limited by synchrotron cooling in the shock we find a very similar estimate $E_{\rm max} \sim 1~{\rm PeV}$, easily reaching the electron energies found by \cite{espinasse2020}. In reality, the situation is more complex. First, the Hillas energy is hard to reach in relativistic shocks \citep[e.g.][]{bell_cosmic-ray_2018} and the maximum energy is highly sensitive to the ability of cosmic rays to amplify and stretch magnetic fields and so to enable effective scattering \citep[e.g.][]{matthews2020}. Second, the magnetic field is likely to be weaker than our estimate given that we over-predict the radio luminosity and \cite{espinasse2020} estimate $B \sim 0.2~{\rm mG}$, and it is also not clear that the reverse shock would persist long enough to explain the late-time X-ray observations. However, our calculation shows that a reverse shock similar to the one that emerges in our simulation is a feasible site of TeV -- and possibly PeV -- particle acceleration for both electrons and protons/ions, motivating further study and dedicated observations.

Our prediction of potential PeV particle acceleration in discrete jet ejecta from XRBs is particularly interesting given recent results from very high energy (VHE) gamma-ray observatories. Both the {\sl The High-Altitude Water Cherenkov (HAWC)} observatory and {\sl Large High Altitude Air Shower Observatory (LHAASO)} have reported detections of VHE gamma-rays from X-ray binaries. HAWC has found extended VHE gamma-ray emission from both SS 433 \citep{Abeysekara2018} and V4641 Sgr \citep{alfaro2024}, while LHAASO found detected five XRBs in $>100$~TeV gamma-rays \citep{lhaaso2024}. One of these latter five sources is our `case study' MAXIJ1820+070, and, interestingly, the gamma-ray emission is offset along the direction of the receding jet component. Our work suggests that particle acceleration in the reverse (and potentially also forward) shock of the discrete jet ejecta provides a plausible mechanism to accelerate particles to the VHE regime. Our findings also strengthen the idea that X-ray binaries may be important sources of cosmic rays up to, and possibly beyond, the knee in the cosmic ray spectrum \citep{heinz_cr_2002,fender2005energization,cooper2020}.

This simulation demonstrates the natural ability for these ejecta to re-accelerate particles late in their deceleration through a long-lived reverse shock. The lifetime of the reverse shock in this simulation is roughly equal to the time between the inferred ejection date and the date at which we begin simulating the ejecta. This means that if the ejecta were launched with this size and energy, the reverse shock would presumably have already passed through the blob within the 90 days. The ejecta at launch, however, are surely more compact and higher in kinetic energy, affecting the lifetime of the reverse shock.

The lifetime of the reverse shock is intrinsically linked with the lightcurve produced, as outlined in section \ref{instabilities_shocks}. Although the reverse shock is confined to a relatively small volume in the simulation, the luminosity at the peak of the simulation lightcurve is an order of magnitude larger than in observations at this point in the blob's trajectory. We note however that our treatment of particle acceleration is simplistic -- it is unlikely that acceleration efficiency is constant everywhere, and requires a more elaborate treatment which will be addressed in future work.  

\subsubsection{The forward shock as an energy interface}

Along with a strong reverse shock, there is also a powerful forward shock which persists throughout the simulation time. In our treatment, only the blob material (i.e. the reverse shock) radiates, an assumption partially motivated by the expectation that stronger magnetic fields would be entrained by the blob and therefore provide a more efficient environment for shock acceleration as compared to the shocked ISM.  It is likely that the forward shock also contributes to particle acceleration, but requires sufficient magnetic field strength in the ISM, and/or significant amplification of such field by the shock itself in order to contribute significantly. In future, discriminating between forward and reverse shock particle acceleration may be possible from combined kinematic and radiation modelling by adapting tools such as {\texttt{jetsimpy}} \citep{wang2024}. In fact, this was pursued in \citet{cooper2025} in the context of the decelerating jets from the XRB MAXI J1535-571, and they concluded (among other things) that early emission was dominated by the reverse shock, whereas the forward shock contributed more to late-time emission.

Despite not contributing to radiation or particle acceleration in this simulation, the forward shock efficiently accelerates and thermalises the ISM. From Fig.~\ref{fig:energetics} it is clear that the ISM is being heated at a faster rate than the blob, and it is the forward shock that ultimately facilitates this heating as it is passing through ISM material. It is difficult to disentangle the role of the forward and reverse shock in this energy transfer, as the two are intrinsically linked energetically. However, the forward shock clearly has an important role in a volumetric sense, allowing the energy input to be spread over a large extent of ISM. Overall, the total energy transfer from blob to ISM as the ejecta decelerate demonstrates that these systems are able to efficiently convert the kinetic energy, obtained close to the compact object through some launching mechanism, into mainly thermal energy in the ISM. It is clear that XRB jets, at least in the `transient ejecta' mode, have the ability to contribute significantly to feedback on the surrounding environment.

\subsection{Comparison with existing models}\label{sec:model_comparison}

The complex structures emergent at later times, as seen in this simulation, are usually not considered when modelling the evolution of the morphology, dynamics, or re-brightening of the ejecta from XRBs. Historically, the ejecta are often modelled as spherical adiabatically expanding plasma blobs which do not interact with the surrounding environment, as originally proposed by \cite{vanderlaan1966}. This model is used to describe the radio spectral evolution of a blob which begins as compact and optically thick, and becomes optically thin at a later point in the expansion. Although this reproduces some characteristics of some radio lightcurves, it is evident that this is not an adequate physical description of the system (see e.g. discussion in \citet{fender2023comprehensive}).

At moderate Lorentz factors -- present in this simulation and likely XRB ejecta in general -- a spherical morphology is not stable on lightcurve evolution timescales, evident particularly in Figure \ref{fig:disruption}. Furthermore, the interaction between the blob and its surrounding environment has a significant (and likely governing) impact on its deceleration and re-brightening, ie. in-situ particle acceleration. In fact, we find in our simulations that the eventual disruption of the blob and the transfer of energy from kinetic to radiative is driven by shocks and instabilities. Corrections to the \cite{vanderlaan1966} model have been made to by \cite{tetarenko2017extreme} to account for geometric/projection effects, and relativistic beaming. In a similar fashion, the `boosted fireball' model by \citet{duffell2013boosted} describes a spherically expanding fireball propagating with a bulk Lorentz factor to produce a beamed jet-like flow -- this is useful for modelling GRB emission, and possibly in a certain regime XRB ejecta as well. However, despite these improvements, the original assumption of the models, ie. non-interacting spherical expansion, remains an oversimplification for these systems both for dynamics and lightcurve evolution in the context of XRBs.

Another model commonly used to describe the \textit{dynamical} evolution of mildly relativistic jets in XRBs (although, again, originally intended for GRBs) is the external shock model developed by \cite{wang2003external}, which has successfully been used to replicate the trajectory of several ejection events (MAXI J1348-630 by \citealt{carotenuto2022modelling} and \citealt{zdziarski2023}, MAXI J1535-571 and XTE J1752-233 by \citealt{carotenuto2024constraining}, XTE J1550–564 by \citealt{wang2003external}, \citealt{hao2009large} and \citealt{steiner2012modeling}, and H1743–322 by \citealt{steiner2011distance} ). The model describes relativistic, symmetric, conical jets with an initial Lorentz factor and kinetic energy -- as the jets expand along their opening angle they sweep up mass and in turn dissipate their kinetic energy through shocks. It is assumed that radiation losses are negligible and expansion is adiabatic. Overall, this seems to provide a more physically motivated model as compared to \cite{vanderlaan1966}, although does not take into account instabilities and thereby non-conical morphological evolution, which, as shown by the 1D model, do play a role in additional deceleration. 

\begin{figure}
    \centering
    \includegraphics[width=1\columnwidth]{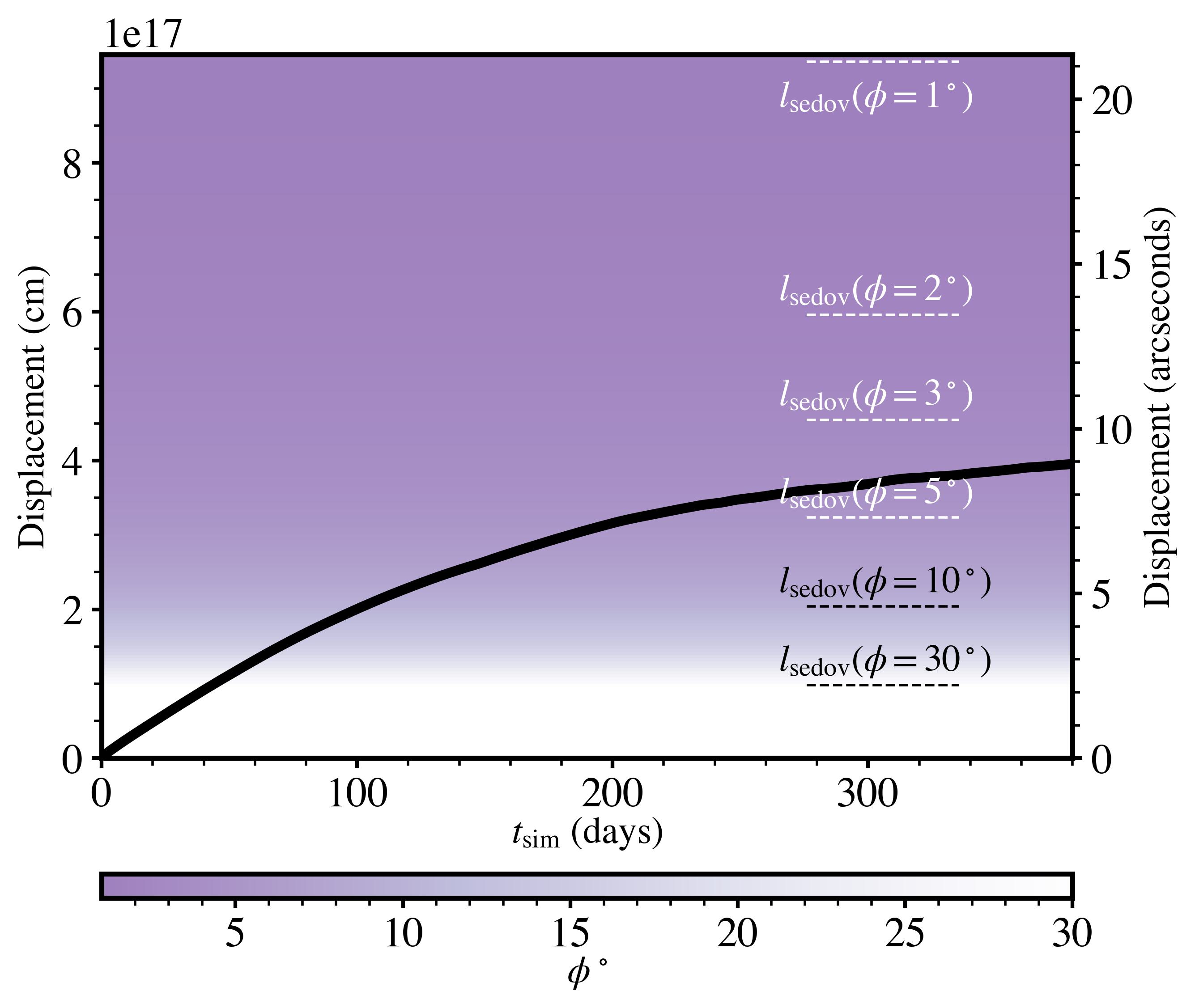}
    \caption{Predictions of the Sedov length $l_{\rm sedov}$ at various half-opening angles, and $E_0$, $\Gamma_0$, and $M_0$ from simulation initial conditions listed in Table \ref{Tab:table_fid}. The deceleration of the centre of emission of the simulation (no geometric viewing effects) is seen as a solid black line, and the gradient represents the continuous half-opening angles as a function of Sedov length.}
    \label{fig:model_comparison}
\end{figure}

The external shock model is formulated around the energy conservation equation
\begin{equation}
    E_0=(\Gamma-1)M_0c^2+\sigma(\Gamma_{\rm sh}^2 - 1)m_{\rm sw}c^2
\end{equation}
where $E_0$ is the initial total kinetic energy in the ejecta, $\Gamma$ is the instantaneous bulk Lorentz factor of the ejecta, $M_0$ is the ejecta mass, $\Gamma_{\rm sh}$ the bulk Lorentz factor of the shocked ISM, $\sigma$ an order unity factor dependent on the velocity regime, and finally $m_{\rm sw}$ the mass of the entrained material. The first term on the right of this equation encapsulates the instantaneous kinetic energy in the ejecta and the second term the internal energy of the ISM as it is swept up. This model predicts the proper motion of the jet with a set of parameters, but the degeneracy between $E_0$, $\phi$, and $n_{\rm ism}$ means that only an `effective' energy can be constrained. In \citet{carotenuto2022modelling} this effective energy is defined as
\begin{equation}
    \bar{E}_0 = E_0\left(\frac{n_{\rm ism}}{ \text{cm}^{-3}}\right)^{-1}\left(\frac{\phi}{1^{\circ}}\right)^{-2}
\end{equation}
where $\phi$ is the half-opening angle of the jet. 

In \citet{carotenuto2024constraining}, they model the angular separation of the ejecta from MAXI J1820 presented in \citet{Bright2020} (discussed in this work), and measure $\bar{E}_0=2.5\times10^{46}$ erg. If we impose conditions from our fiducial simulation, $n_{\rm ism}=10^{-3}$  and $E_0=2.16\times10^{44}$ erg, this necessitates a half-opening angle $\phi\sim3^{\circ}$.  In Figure \ref{fig:multipanel-sim}, if one estimated the width of the brightest section of emissivity at $t_{\rm sim}\sim300$ days as approximately 1 arcsec, at roughly 14 arcsec from the core this results in an estimated half-opening angle $\phi_{\rm sim}\approx2^{\circ}$, consistent with the model results. However, as is also apparent in this Figure, the conical jet structure with a well-defined opening angle is not necessarily an accurate description of the structure of the jet ejecta. We note that \citet{tetarenko2021measuring} measure $\phi\approx0.4^\circ$ from the emission and spectral characteristic of the compact jets in the source, a much narrower opening angle than what we derive for these ejecta.  

Importantly, these simulations allow us to begin eroding degeneracy in effective energy with $\phi$ and $n_{\rm ism}$. Firstly, $\phi$ is not a fixed parameter per se, in that it is a results of initial conditions with some hydrodynamical evolution, and can be estimated directly from simulations. Secondly, we can at least put limits on $n_{ism}$ and $E_0$ from energetic arguments, explored in more detail in section \ref{accretion-jet-connection}, and demonstrated in \citet{carotenuto2022modelling}. In particular, with better modelling of particle acceleration, the lightcurve will help disentangle $E_0$ irrespective of $n_{\rm ism}$.

The external shock model exists within the larger framework of the standard forward-reverse shock model for GRBs (\citealt{rees1992relativistic},  \citealt{piran1995hydrodynamic} \citealt{kobayashi2000light}, \citealt{zhang2022semi}). Aspects of this model can be modified for the mildly relativistic regime and used to describe XRB phenomena -- the extent to which they are more widely applicable to XRBs has been investigated in more detail by \cite{matthews2025}. The dynamical evolution of the shells ejected in GRBs is commonly signposted by characteristic radii (equivalent to propagation distances, but in spherical or quasi-spherical geometry), the most ubiquitous of which is the Sedov length $l_{\rm sedov}$:  the radius where the material swept up by the ejecta has accumulated rest mass energy equal to the initial energy, and the characteristic length-scale for the interaction between a relativistic flow and it's external medium. It is defined as:
\begin{equation}
    l_{\rm sedov}= \left(\frac{3 E_0}{\rho_{\rm ism} c^2 \pi\phi^2}\right)^{1/3}=
    \left(\frac{3(\Gamma_0-1)M_0}{\rho_{\rm ism} \pi\phi^2}\right)^{1/3}
\end{equation}
where $\Gamma_0$ and $M_0$ are the bulk Lorentz factor and rest mass of the ejecta at launch, $E_0=(\Gamma_0-1) M_0 c^2$ for kinetically dominated ejecta, and we have assumed small opening angles such that the solid angle of the conical section is approximated by $\Omega \approx \pi \phi^2$. Although this conical geometry is not a perfect equivalent to the geometry of our simulation, it can still be a useful comparison.

We calculate $l_{\rm sedov}$ with the initial conditions of the fiducial simulation for various values of $\phi$ as seen in Figure \ref{fig:model_comparison}, and we indicate in the figure where this corresponds to with respect to the displacement of the centre of emission in the simulation (which we remind the reader is very similar to what is directly observed). One would expect that significant deceleration would be observable around $l_{\rm sedov}$, as the deceleration is primarily driven by energy transfer between the blob and the ISM (as made clear in Section \ref{energetics}). In the fiducial simulation, $l_{\rm sedov}$ is consistent for half-opening angles of about $\phi\sim4^\circ$, roughly consistent with the external shock model. Overall, our simulations align with basic predictions from GRB models, and even produce comparable phenomena such as a forward and reverse shock structure. 


We note that $E_0$  and $\Gamma_0$ values are obtained from the simulation at $t_{\rm sim}=0$ days, thus $t_{\rm obs}=90$ days, therefore one would expect these to be lower limits on the values at launch. In addition, these values are obtained from a simulation which is not a direct fit to the data, but instead gives us an idea of the possible and likely parameter space given that simulations in this regime reproduce the data well.


\subsection{Cavitation of the ISM}\label{cavitation}

In the simulation, we observe a distinct high-pressure, low-density cavity being blown up around and in the wake of the blob as it propagates. This is important for two main reasons: firstly, because it provides an explanation for the existence of a low-density environment necessary for the propagation of the ejecta to observed distances; secondly, it provides a mechanism for the formation of an abrupt cavity `wall' or density jump as inferred by observations of other decelerating ejecta. 

In order to match the observed deceleration without an unrealistically high kinetic energy, we require a very low-density environment (see Sections \ref{fidsim} and \ref{decel_section}). This balance between density and energy has also been explored in the aforementioned 1D blast-wave modelling, which constrains the parameter $E_0 / \phi^2 n_{\rm ism}$ and makes explicit this degeneracy \citep{heinz2002radio-lobe,wang2003,steiner2012modeling,carotenuto2022modelling,carotenuto2024constraining}. This trade-off is an intuitive expectation -- more energy is required to clear out more mass in the path of the projectile.

These low density environments could correspond to a preferred, albeit somewhat extreme, (hot coronal) phase of the ISM, or be explicitly formed by external mechanisms. Our simulations suggests one such factor: that previous transient jet activity can itself be responsible for creating the low density environment needed. Low-mass X-ray binaries cycle through repeated outbursts, so relics of previous jet activity might be naturally expected. The exact environment would presumably depend on a complex interplay between the outburst duty cycle, jet energetics, refilling timescales, and underlying structure of the local ISM. We can, however, roughly estimate the longevity of the cavity created in this simulation by first estimating the minimum timescale after which the cavity would come into pressure equilibrium with the ISM. Taking the lateral cavity width, speed of the shock, and pressure within the cavity at the end of the simulation time ($\sim 350$ days, noting that the simulation begins at $t_{\rm obs}=90$ days and thus we would expect the cavity to be larger in reality), the time for the bubble expanding at constant speed to decompress to ISM pressure is $\tau_{\rm expand}\sim16$ years, at which point the final radius is $R_{\rm expand}\sim 2\times10^{18}$ cm (roughly 10x the radius at 350 days). After this point, the refilling time can be calculated as $\tau_{\rm refill}\sim R_{\rm expand}/c_{s}$ where $c_s$  is the estimated sound speed within the expanded and equilibrated cavity with density equal to the average density in the cavity at the end of the simulation, giving $\tau_{\rm refill}\approx 8\times10^{10}$ seconds, $\sim 2500$ years. The sum of these two conservatively estimated timescales is well above plausible outburst recurrence timescales, which in observed XRBs are years to centuries \citep{tanaka1996,mcclintock2006}.

Indeed, the idea that XRBs exist inside under-dense cavities has been used to explain deceleration behaviour of other transient ejecta (e.g. \cite{sikora2023}). A particularly notable LMXRB source which also exhibits behaviour suggesting the existence of a locally under-dense ISM cavity is MAXI 1348-630, as studied by \cite{carotenuto2022modelling} and \cite{zdziarski2023}. It appears that number densities $\ll 1$ cm$^{-3}$ are required in the region surrounding the binary system in order to match the observed propagation of the transient ejecta launched in the 2019 outburst. \cite{carotenuto2022modelling} propose the existence of a cavity `wall' which the ejecta impact and decelerate accordingly, although this model requires kinetic energy in excess of the estimated accretion power. \cite{zdziarski2023} remodel the data with an additional `transition region' between the cavity and cavity wall, where the density grows exponentially, which allows for a more physically motivated kinetic energy. The cavity density derived for the best-fitting transition region model is $n_{\rm cavity}\approx 10^{-5}$ cm$^{-3}$ , and that for the cavity wall model is $n_{\rm cavity}\approx 10^{-3}$cm$^{-3}$. In our results, we propagate the ejecta into `pristine' ISM with density akin to the cavity modelled by \cite{carotenuto2022modelling} ($n_{\rm ism}=10^{-3}$cm$^{-3}$), and the blob itself then creates a cavity an order of magnitude lower density than the surrounding medium -- this is in contrast to the cavities in \cite{carotenuto2022modelling} and \cite{zdziarski2023} which are 3-5 orders of magnitude (respectively) less dense than their surrounding pristine medium which is modelled with a number density of 1 cm$^{-3}$. It is possible that through multiple cycles of jet activity a deeper cavity could be excavated, which could replicate the density jump suggested by this modelling of the MAXI J1348-630 observations. 

We also observe that the lateral edges of the cavity in the simulation do roughly resemble the transition region described by \cite{zdziarski2023} with an exponential increase in density. However, the edge is shocked and therefore creates a sharp peak in density before returning back to the ambient density: this is particularly apparent in Figure \ref{fig:multipanel-sim}, leftmost panel in the bottom row (density at late-time), where the cavity is less dense than the surrounding medium, but the densest region is at the shock front. This could motivate more detailed models of the cavity edge, but the density profile at the head of the jet is much more turbulent and complex, demonstrating that simple models may fail in regions that fall in the direct wake of previous ejecta. 

The estimated maximum size of the bubble after expansion is much larger than propagation distances typically observed, so it is also reasonable to expect that many ejecta do not encounter this 'wall' and instead propagate to deceleration/destruction without impacting significant density features. We infer that this is likely the case for MAXI J1820+070 (and two of the other sources modelled by \cite{carotenuto2024constraining}) given that the smooth deceleration profile is reproduced through propagation in a homogeneous ISM. Perhaps observations of abruptly decelerating ejecta, such as in MAXI J1348-630, are due to relatively short recurrence times for jet activity and/or variation in the energetics of transient ejecta. 

Overall, the densities in the simulation paint the picture that the history of ejections can create locally varying density environments -- and especially {\em low} density environments -- which are then reflected in the propagation behaviour of later ejections. 




\subsection{Kinetic energy and jet power}\label{accretion-jet-connection}
At 90 days post-launch, the ejected blob in the simulation requires a substantial amount of kinetic energy in order to propagate far enough to match the observed deceleration. This necessitates both a significant amount of mass within the blob itself and a moderately high Lorentz factor. Given a kinetic energy of $E_{\rm K}=2\times10^{44}$ erg and an internal energy of $E_{\rm int}=2\times10^{43}$ erg, we can estimate the power required of the symmetric jets during launch as $P_{\rm jet}=2E_{\rm tot}/\Delta t$, where $\Delta t$ is the time over which the jet is launched, and $E_{\rm tot}=E_{\rm K}+E_{\rm int}\approx E_{\rm K}$ if the jet is kinetically dominated. Since the radio flare is coincident with the supposed launching of the jet, it is possible that the rise-time of the flare might provide a timescale over which the jet is launched.  \cite{Bright2020} observe a rise-time of $\Delta t \sim 6.7$ hours, or 4.5 hours in the rest-frame, requiring a jet power of $P_{\rm jet}=3 \times 10^{40}$ erg s$^{-1}$. It is also important to note that our kinetic energy estimate provided here is for 90 days post-launch; the kinetic energy is likely even larger at $t=0$.  

This high jet power can pose several problems for accretion and jet ejection mechanisms as we currently understand them. Firstly, it appears that the jet power exceeds the maximum value of the X-ray bolometric luminosity. With an X-ray luminosity peak during the flaring episode of about $L_{1-10\rm keV}=1.4\times10^{38}$ erg s$^{-1}$ \citep{Bright2020}, we estimate the bolometric luminosity as $L_{\rm bol}=\xi L_{\rm 2-10keV}$ where $\xi=0.8$ for BHs in X-ray outburst \citep{migliari2006jets} and therefore $L_{\rm bol}\approx2\times10^{38}$erg s$^{-1}$ , corresponding to about 16\% of the Eddington luminosity for a BH of this mass. Assuming that the radiative efficiency $\eta_{\rm rad}$ of the BH is $\sim10\%$, and that it is related to the mass accretion rate through $\eta_{\rm rad}\dot{M}c^2=L_{\rm bol}$, we are able to estimate a mass accretion rate of $\dot{M}=2.2\times10^{18}$g s$^{-1}$, or $\dot{M}c^2=2\times10^{39}$erg s$^{-1}$. Given that $P_{\rm jet}\gg\dot{M}c^2$, there is evidently a conundrum of some sort. There are several avenues one can consider in order to resolve this energy discrepancy, not all of which are necessarily independent, and which may all be contributing in some capacity.\footnote{We also refer the reader to section 5.4 of \cite{carotenuto2024constraining} for a similar relevant discussion.} 

\begin{enumerate}
\item {\bf Energy and ISM density:} Firstly, it is not unlikely that the kinetic energy is over-estimated. In our prescription, we choose not to lower the ISM density beyond the lowest density phase (a hot coronal ISM) but it is possible, as discussed in the previous section, that the ISM could be lower density still, possibly due to previous ejections or even the hard-state jet (although our investigation only evidences the former and not the latter). This lower density could allow for a reduced kinetic energy while maintaining relativistic velocity by instead lowering the mass density of the blob. For example, a subsequent ejection would propagate through an ISM roughly 10 times less dense than the current ISM estimation thanks to the cavity created by the primary blob (our simulation), assuming that the second blob is launched within some reasonable timescale over which the cavity does not refill. There is no sign of refilling within the simulation time ($\sim350$ days), and the minimum timescale over which we estimate the cavity to persist is about 2.5 thousand years (detailed in section \ref{cavitation}) -- plenty of time for the possibility of a second ejection cycle. Maintaining the density contrast $\chi$, which roughly determines the propagation distance for a fixed velocity and radius, we would be able to reduce the kinetic energy (through reduction in blob density) by an order of magnitude and therefore reduce the jet power to $\approx\dot{M}c^2$. If what we are seeing is indeed a subsequent ejection, the prior ejection or ejecta went undetected, not unlikely given the birth of X-ray astronomy only $\sim80$ years since these data. Additionally, the lower density of the blob itself would bring it further within the back-of-the-envelope mass constraints from the accretion rate. We note here two things regarding the maximum mass estimate: firstly, this does not account for the possibility of entrainment, and it is feasible that the jet was initially very fast and light, and subsequently swept up mass as it decelerated therefore allowing for heavier ejecta; secondly, the maximum mass calculation assumes that all accreted rest-mass can be ejected in the jet, which inherently can not be true for a radiating system unless energy is being extracted from the spin of the black hole.

\item {\bf Timescale:} The next most obvious offender in this power discrepancy is the estimation of the timescale over which the jet is launched, $\Delta t$. In order to reconcile $P_{\rm jet}$ with the estimate of $\dot{M}c^2$, the timescale would have to extend to 3.6 days in the rest frame, and 5.4 days in the observer frame. These timescales are calculated assuming the paradigm where the maximum jet power is $P_{\rm jet}\approx1.3a_*^2\dot{M}c^2$ where $a_*$ is the BH spin, and this maximal jet power equation arises from the numerical modelling of a Magnetically Arrested Disk (MAD) by \cite{tchekhovskoy2015formation} simulating the extraction of energy via the black hole spin, ie. the Bladford-Znajek process \citep{blandford-znajek1997}. This is not a strict theoretical upper limit for jet power, but is likely in the right ballpark for spin-powered jets. The upper limit for jet power assuming the MAD state at this particular accretion rate is therefore $P_{\rm jet}\approx 2.6\times10^{39}$ erg s$^{-1}$. This rest-frame 3.6 day timescale is significantly longer than the 4.5 hour rest-frame rise time of the flare. 

Other prescriptions for measuring the launch timescale of transient jets have been proposed, mainly linked with X-ray timing properties prior to the radio flaring, and specifically the detection of type-B quasi-periodic oscillations (QPOs) in the X-ray power spectrum thought to be produced by the precession of the jet. In particular, it has been hypothesized that the switch between type-C (also likely linked to precession, but of the inner flow) and type-B QPOs signifies the beginning of the launch of the transient ejecta (\citealt{russell2019disk}, \citealt{fender2009jets}, \citealt{miller2012disc}). A very strong type-B QPO was observed in MAXI J1820+070 by \cite{homan2020rapid} just prior to the radio flare associated with the launching of the ejecta simulated in this paper, and the duration of this oscillation phase could provide a more accurate estimate of the launching timescale. However, the QPO was only present for 2-2.5 hours prior to the flare period, and thus only further constrains the launching timescale (and correspondingly increases the jet power requirement). Additionally, the LMXRB GRS 1915+105 as observed by \citet{fender2000giant}, exhibited rapid, large amplitude repeated ejections where the duration between discrete ejection events was $\sim 20$ minutes. Mounting evidence to suggests that launching timescales tend to be on the order of minutes to hours, rendering a 3.6 days timescale unlikely. 

\item {\bf Magnetic fields:} It is also possible that by including large-scale magnetic fields in the simulation, we would be able to reduce the kinetic energy and maintain a similar length trajectory. The magnetic fields could provide a barrier for the disruption of the blob, as has been demonstrated in studies of cloud-crushing \citep{jones1996,mccourt2015,jung2023}. Slower disruption would lead to a larger propagation distance (for fixed energy) and thus a lower energy requirement, clearly motivating future work using special relativistic MHD simulations. 

\end{enumerate}


Similar arguments to the above can be made in terms of mass conservation, and indeed we used this to inform our initial blob density. However, the mass constraint is less fundamental than the energy constraint, because mass-loading may come primarily from entrainment, possibly the sweeping up of the pre-existing jet. It has even been proposed that the initial flaring episode could be a direct result of the interaction between the transient jet launched and the previous hard-state jet \citep{fender2004towards}. This remedies the issue of taking mass directly from the accretion flow, as in the maximum mass estimate we unrealistically assume that no mass is advected by the BH, and also allows for a powerful jet from a radiatively efficient flow. Nonetheless, as summarized previously, the maximum jet power, from the \citet{tchekhovskoy2015formation} model, is still at least tenfold too small to account for the calculated jet power.

Overall, it is clear that a high kinetic energy and the resulting high power requirements of the transient jet challenge our understanding of accretion and ejection mechanisms in the regions close to the black hole. It is possible to lower the energy requirements by considering 
lower density ISM, a longer jet launching timescale, or dynamically important magnetic fields in the ejecta. 
We note that jets of this high power are not unique to this system and have been inferred even in the very first superluminal source detected, GRS 1915+105 \citep{mirabel1994superluminal}. MAXI J1348-630 \citep{carotenuto2022modelling}, XTE J1550-564 \citep{steiner2012modeling} have also produced jets with comparably high kinetic energy estimates.




\section{Conclusions}\label{limitations_future_work}
\label{conclusions}


The detailed campaign of radio observations following the flaring detected from MAXI J1820+070 produced a rich dataset mapping out the kinematic and radiative evolution of one of the most long-lived discrete ejections observed from a stellar mass black hole. We exploit some of this data, through the use of hydrodynamic simulations and careful reconstruction of realistic radio images from these simulations, to uncover new insights, and new intuition, about these strange phenomena. 

We see from our simulations that the ejected blob, 90 days after launch, begins with more kinetic than internal energy, and transfers this energy into the ISM as it decelerates. Most of the deceleration is mediated by the transfer of blob kinetic energy into ISM thermal energy at the forward shock. In essence, these ejecta act as an efficient mechanism to take energy from close to the horizon of the BH, where the jet is launched, and efficiently deposit this energy into the ISM on parsec scales, providing a means of stellar mass BH feedback. 

A minority of the initial kinetic energy is also converted into thermal energy in the blob by the passage of the reverse shock through the ejecta, contributing to observable luminosity and potentially in-situ particle acceleration (feasibly up to TeV to PeV energies). 


For the blob to propagate to the distance observed by \citet{Bright2020}, the surrounding ISM must be sufficiently tenuous and the blob sufficiently dense in comparison. We also observe in simulations the formation of a long-lived under-dense cavity which provides a mechanism for forming these low-density environments --  we propose that the already low ISM density could be the result of a previously undetected ejection clearing out the ISM in its wake. The existence of these cavities is supported by the observational signatures of significant density jumps in the ISM profile surrounding some XRBs.

We are able to replicated the deceleration of the observed ejecta from 90 days post-launch onwards. The blob in the simulation requires a significant amount of kinetic energy -- if the ejecta are launched over the same timescale as the coincident flaring of the core, this also sets a very large power requirement for the jets, where $P_{\rm jet}\gg\dot{M}c^2$. This is not the first instance where measured XRB jet powers exceed first-principle expectations. The jet power, however, can be lowered through several mechanisms: most likely by lowering the ISM density (allowing for a lower kinetic energy), and possibly by the incorporation of stabilizing large-scale magnetic fields in the ejecta. 

In addition to modelling the kinematics of the ejecta, we also predict realistic radio images at several key points in the evolution. We successfully reproduce the MeerKAT radio images at $t_{\rm obs}=90$ days, but do not recover the same flux or morphology for the simultaneous eMERLIN predictions. This is due to simplifying assumptions made about size constraints, and the complex relationship between source morphology and $u$,$v$ coverage of the radio telescopes. Despite not reproducing the images, we are able to improve previous constraints on not only the size but the flux distribution of the ejecta. We demonstrate that the simulation-to-radio pipeline \texttt{BLOB-RENDER} can improve our understanding of both simulation results and observations. 

Overall, much can be gained from relatively simple simulations if they are well informed by observations, and can be directly compared back to such observations.

\section*{Acknowledgements}

The authors thank the anonymous referee for their insightful comments, which improved the quality of the manuscript. KS acknowledges support from the Clarendon Scholarship Program at the University of Oxford and the Lester B. Pearson Studentship at St John's College, Oxford. JM acknowledges funding from a Royal Society University Research Fellowship (URF$\backslash$R1$\backslash$221062). IH acknowledges support of the Science and Technology Facilities Council (STFC) grants [ST/S000488/1] and [ST/W000903/1] and from a UKRI Frontiers Research Grant [EP/X026639/1], which was selected by the European Research Council. RF acknowledges support from UKRI, the European Research Council and the Hintze family charitable foundation. IH acknowledges support from the South African Radio Astronomy Observatory which is a facility of the National Research Foundation (NRF), an agency of the Department of Science and Innovation. IH acknowledges support from Breakthrough Listen. Breakthrough Listen is managed by the Breakthrough Initiatives, sponsored by the Breakthrough Prize Foundation.

We gratefully acknowledge the use of the following software packages: \textsc{pluto} \citep{mignone_pluto_2007}, matplotlib \citep{Hunter:2007}, VisIt \cite{HPV:VisIt} used to make Figure \ref{fig:3dsim},  and astropy:\footnote{\url{http://www.astropy.org}} \citep{astropy:2013, astropy:2018, astropy:2022}. We thank and acknowledge access to eMERLIN data through David Williams-Baldwin (who was also particularly helpful with regards to the data), and MeerKAT data through the ThunderKAT collaboration.

The authors would like to acknowledge the use of the University of Oxford Advanced Research Computing (ARC) facility in carrying out this work (\url{http://dx.doi.org/10.5281/zenodo.22558}).

Lastly, we would like to acknowledge collaborators Alex Cooper, Francesco Carotenuto and Joe Bright for productive and helpful discussions.

\section*{Data Availability}

The data underlying this article are available from the authors on reasonable request.  


\bibliographystyle{mnras}
\bibliography{references,JM_refs} 




\appendix


\bsp	
\label{lastpage}
\end{document}